\documentclass[aps,prd,twocolumn,groupedaddress,floatfix,showpacs,preprintnumbers]{revtex4}

\usepackage{graphicx}

\newcommand{\lwig}{\mbox{\,\raisebox{.3ex}
    {$<$}$\!\!\!\!\!$\raisebox{-.9ex}{$\sim$}\,}}
\newcommand{\gwig}{\mbox{\,\raisebox{.3ex}
    {$>$}$\!\!\!\!\!$\raisebox{-.9ex}{$\sim$}}\,}

\newcommand{\bwide}{\begin{widetext}}
\newcommand{\ewide}{\end{widetext}}
\newcommand{\beq}[1]{\begin{equation} \label{(#1)}}
\newcommand{\eeq}{\end{equation}}
\newcommand{\bea}[1]{\begin{eqnarray} \label{(#1)}}
\newcommand{\eea}{\end{eqnarray}}
\newcommand{\nn}{\nonumber}
\newcommand{\rf}[1]{(\ref{(#1)})}
\def\half{\frac{1}{2}}

\def\third{\frac{1}{3}}
\def\nue{\nu_e}
\def\numu{\nu_\mu}
\def\nutau{\nu_\tau}

\begin{document}

\preprint{DESY 03-219 \hspace{6ex} SLAC-PUB-10302}

\title{Relic Neutrino Absorption Spectroscopy}


\author{Birgit~Eberle}
\author{Andreas~Ringwald}
\affiliation{Deutsches Elektronen-Synchrotron DESY, Notkestr. 85, D-22607
Hamburg, Germany}
\author{Liguo~Song}
\affiliation{Department of Physics \& Astronomy, Vanderbilt University, Nashville, TN 37235, USA}
\author{Thomas~J.~Weiler}
\affiliation{Department of Physics \& Astronomy, Vanderbilt University, Nashville, TN 37235, USA}
\affiliation{Kavli Institute for Particle Astrophysics and Cosmology,
        Stanford Linear Accelerator Center, 
        2575 Sand Hill Road, Menlo Park, CA 94025, USA}



\begin{abstract}
Resonant annihilation of extremely high-energy cosmic neutrinos on 
big-bang relic anti-neutrinos (and vice versa) into Z-bosons 
leads to sizable absorption dips in the neutrino flux to be 
observed at Earth. The high-energy edges of these dips are fixed, via the resonance energies, 
by the neutrino masses alone.
Their depths are determined by the cosmic neutrino background density, 
by the cosmological parameters determining the expansion rate of the universe,
and by the large redshift history of the cosmic neutrino sources.
We investigate the possibility of determining the 
existence of the cosmic neutrino background within the next decade from a measurement 
of these absorption dips in the neutrino flux. As a by-product, we study 
the prospects to infer the absolute neutrino mass scale. We find that, with 
the presently planned neutrino detectors (ANITA, Auger, EUSO, OWL, RICE, and SalSA) 
operating in the relevant energy regime above $10^{21}$~eV, 
relic neutrino absorption spectroscopy becomes a realistic possibility. 
It requires, however, the existence of extremely powerful
neutrino sources, which should be opaque to nucleons and high-energy photons
to evade present constraints. 
Furthermore, the neutrino mass spectrum must be quasi-degenerate to optimize the dip, 
which implies $m_\nu\gwig\, 0.1$~eV for the lightest neutrino. 
With a second generation of neutrino detectors, 
these demanding requirements can be relaxed considerably.  
\end{abstract}

\pacs{14.60.Pq, 98.70.Sa, 95.85.Ry, 95.35.+d}

\maketitle

\section{Introduction}

Neutrinos are the elementary particles with the weakest known interactions. 
Correspondingly, they can propagate to us through the cosmic microwave and neutrino 
background (CMB and C$\nu$B, respectively) 
without significant energy loss even from cosmological distances. 
A possible exception to this transparency 
is resonant annihilation of extremely high energy cosmic neutrinos 
(EHEC$\nu$'s) on big-bang relic anti-neutrinos (and vice versa) into 
Z-bosons~\cite{Weiler:1982qy,Weiler:1983xx,Roulet:1992pz,Yoshida:1996ie}. 
This occurs near the respective resonance 
energies,
\begin{eqnarray}
\label{Eres}
E_{\nu_i}^{\rm res} = \frac{m_Z^2}{2\,m_{\nu_i}} = 4.2\times 10^{22}\ {\rm eV}  
\left( \frac{0.1\ {\rm eV}}{m_{\nu_i}}\right)
\,, 
\end{eqnarray}
with $m_Z=91.2$~GeV denoting the mass of the Z-boson~\cite{Hagiwara:fs}
and $m_{\nu_i}$ ($i=1,2,3$) the non-zero neutrino masses 
-- for which there is rather convincing evidence inferred from the apparent observation of neutrino 
oscillations~\cite{Hagiwara:fs}. On resonance, the corresponding cross-sections are enhanced
by several orders of magnitude.  This leads to 
a few percent probability of annihilation within the Hubble radius of the universe, 
even if one neglects further enhancing effects due to cosmic evolution. 
Indeed, it appears that 
-- apart from the indirect evidence 
to be gained from cosmology, e.g., big-bang nucleosynthesis and large-scale structure 
formation -- this annihilation mechanism 
is the unique process~\footnote{For earlier and related suggestions, see 
Ref.~\cite{Bernstein:1963} and Ref.~\cite{Wigmans:2002rb}, respectively. 
For a recent review, also covering non-cosmic-ray opportunities to detect the 
relic neutrinos, see Ref.~\cite{Ringwald:2003qa}.} 
having sensitivity to the C$\nu$B~\cite{Weiler:1982qy}.
Moreover, observation of the absorption dips would present one of the few opportunities to 
determine absolute neutrino masses~\cite{Paes:2001nd}.          
However, the mechanism requires that there exists a sufficiently large EHEC$\nu$ flux at the 
resonant energies of Eq.~(\ref{Eres}). One of the purposes of this work is to quantify how large 
this  EHEC$\nu$ flux must be.

Since the original proposal in 1982~\cite{Weiler:1982qy}, significant advances
have been made in theoretical and observational cosmology, experimental neutrino physics, 
and EHEC$\nu$ physics. Each of these three areas impacts immediately on the 
C$\nu$B measurement. 
Thus, it is timely to investigate again the possibility of 
determining the existence of the C$\nu$B 
and to study the prospects for determining the neutrino masses via resonant neutrino absorption.

What are the new findings that affect the EHEC$\nu$ absorption probability? 

\begin{itemize}
\item The original calculation of neutrino absorption was done for a matter-dominated flat
universe without a cosmological constant.
Recent observations of large-scale gravity, deep-field
galaxy counts, and Type Ia supernovae favor a universe with energy fractions $\Omega_\Lambda\approx 0.7$
in the cosmological constant and $\Omega_M\approx 0.3$ in 
(mainly cold and dark) matter~\cite{Spergel:2003cb,Tegmark:2003ud}. 
The position of the first Doppler peak in recent CMB measurements strongly suggests
that the universe is flat, i.e. the fractional contribution of curvature energy $\Omega_k$ is
negligibly small. These cosmological parameters, together with the Hubble parameter $H_0$, 
determine the expansion rate of the universe as a function of look-back distance.
This expansion history, in turn, 
crucially impacts the EHEC$\nu$ fluxes at Earth, since their sources 
are almost certainly located at cosmological distances~\footnote{ 
In fact, the depths of the absorption dips are inversely proportional to 
the expansion rate, leading, for fixed $H_0$, to a significant increase in the depth by a 
factor of $\Omega_M^{-1/2}\sim 1.8$ at large redshift, when compared to the old 
cold dark matter (CDM) estimate  
based on $\Omega_M=1$~\cite{Weiler:1982qy,Weiler:1983xx} (see \S~\ref{propagation} for details). 
This is equivalent to an improvement in 
statistics by a factor of $\Omega_M^{-1} \sim 3$.}.

\item The oscillation interpretation of the atmospheric neutrino data offers a lower limit on 
the heaviest of the three mass-eigenstates
of
\begin{equation} 
\label{lim_low_atm}
m_{\nu_3}\geq\sqrt{\triangle m_{\rm atm}^2}> 0.04\ {\rm eV}
\end{equation}
at 95\,\% confidence level (CL), from the inferred mass splitting 
$\triangle m_{\rm atm}^2$~\footnote{We will label neutrino masses in the order 
$m_{\nu_1}\leq m_{\nu_2}\leq m_{\nu_3}$, 
regardless of the type of neutrino mass spectrum (``normal'' or ``inverted''). 
In this convention, $\triangle m_{\rm atm}^2\simeq \triangle m_{32}^2$,
$\triangle m_\odot^2\simeq \triangle m_{21}^2$ in 
a ``normal'' scheme, and 
$\triangle m_{\rm atm}^2\simeq \triangle m_{21}^2$,
$\triangle m_\odot^2\simeq \triangle m_{32}^2$ in an 
``inverted'' one, respectively. (We are using the obvious notation
$\triangle m_{ij}^2\equiv m_{\nu_i}^2-m_{\nu_j}^2$.)   
Another convention is often used in the literature: 
$m_{\nu_1}\leq m_{\nu_2}\leq m_{\nu_3}$ in ``normal'' schemes, 
and $m_{\nu_3}\leq m_{\nu_1}\leq m_{\nu_2}$ in ``inverted'' schemes;
this has the feature that,  
in either scheme, 
$\triangle m_\odot^2\simeq \triangle m_{21}^2\geq 0$
and $\triangle m_{\rm atm}^2\simeq |\triangle m_{32}^2|$.}. 
On the other hand, studies of the cosmic evolution of the large-scale structure, 
as observed today,   
from primordial density perturbations, measured in the CMB,  
yield an upper bound on the sum of 
neutrino masses~\cite{Spergel:2003cb,Hannestad:2003xv,Tegmark:2003ud}
\begin{equation}
\label{lim_cosm}
\sum_i 
m_{\nu_i} \lwig 1.2\ {\rm eV}\,.
\end{equation}
Since oscillation studies also reveal that neutrino mass-splittings are small compared to the
eV scale, the cosmic bound per mass-state is conservatively~\footnote{
Seemingly minor assumptions can make a crucial difference for neutrino inferences
from cosmic structure evolution, as discussed in detail in 
Refs.~\cite{Hannestad:2003xv,Tegmark:2003ud}. In this sense, the recent 
claim that a combination of WMAP, 2dFGRS, and X-ray galaxy cluster data prefer a 
non-zero neutrino mass of $\sum_i m_{\nu_i} = 0.56^{+0.30}_{-0.26}$~eV~\cite{Allen:2003pt} 
have to be viewed with some caution.} $\sim 0.4$~eV -- about a factor
of five better than the laboratory bounds inferred from tritium beta decay~\cite{Barger:1998kz,Weinheimer:tn},  
$m_{\nu_3}<2.2$~eV (95\,\% CL)~\footnote{The neutrino mass parameter measured in tritium
beta decay is 
$m_\beta \equiv (\sum_i \mid U_{ei}\mid^2 m_{\nu_i}^2)^\half$. 
Therefore, the heaviest neutrino mass is bounded by 
$m_{\nu_3}<\sqrt{m_\beta^2+\triangle m_{\rm atm}^2}$~\cite{Barger:1998kz}.
}, 
and neutrinoless double beta decay~\cite{Klapdor-Kleingrothaus:2000sn}, 
$m_{\nu_3}\lwig (0.66\div 2.70)$~eV
~\footnote{This bound applies only for
Majorana neutrinos. For a quasi-degenerate neutrino spectrum 
($m_{\nu_1}\gg \triangle m^2_{\rm atm}$),  
it arises from $m_{\nu_3}\leq 2\,\langle m_\nu\rangle$, where $\langle m_\nu\rangle$ 
is the effective mass  $\mid \sum_i U_{ei}^2\,m_{\nu_i}\mid$, measured
in neutrinoless double beta decay, $U$ being the leptonic mixing matrix. 
The quoted ranges in the upper bound on $\langle m_\nu\rangle$ take into account the spread 
due to different calculations 
of the relevant nuclear matrix elements. 
A recently reported evidence for neutrinoless double beta decay~\cite{Klapdor-Kleingrothaus:2001ke} 
and correspondingly deduced parameter range 
$\langle m_\nu\rangle = 0.39^{+0.17}_{-0.28}$~eV (95\,\% CL) have been 
challenged by Ref.~\cite{Feruglio:2002af}.
Nevertheless, in our conclusions we will note that a quasi-degenerate neutrino mass 
above $\sim 0.1$~eV is required to produce a measurable absorption dip 
at an accessible resonant energy.}.
Thus we have
\begin{equation}
\label{lim_comb_osc_cosm}
0.04\ {\rm eV} < m_{\nu_3}\,\lwig\, 0.4\ {\rm eV}\,. 
\end{equation}
It is remarkable that the neutrino mass, whose value was 
compatible with zero at the time of the original proposal for Z-dips~\cite{Weiler:1982qy}, 
is now known to be not only non-zero,
but to lie within a one-order of magnitude range!
Accordingly, the resonant annihilation energy for the heaviest mass-state is 
also known to within an order of magnitude:
\begin{eqnarray}
\label{Eres_lim}
1\times 10^{22}\ {\rm eV}\,\lwig\, E_{\nu_3}^{\rm res} < 1\times 10^{23}\ {\rm eV}  
\,.
\end{eqnarray}

\begin{figure}
\begin{center}
\includegraphics*[bbllx=20pt,bblly=221pt,bburx=572pt,bbury=608pt,width=8.6cm]{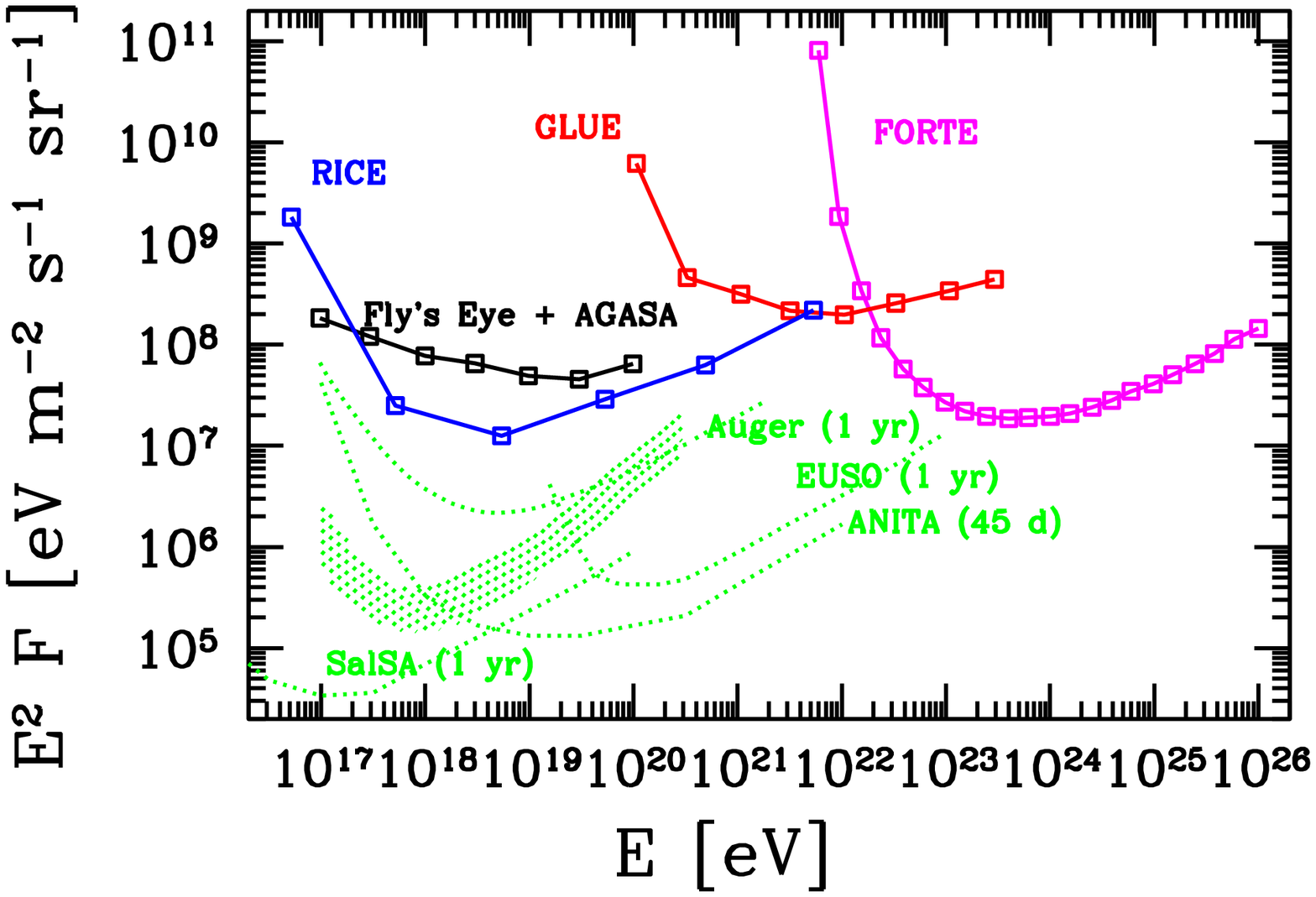}
\includegraphics*[bbllx=20pt,bblly=221pt,bburx=572pt,bbury=608pt,width=8.6cm]{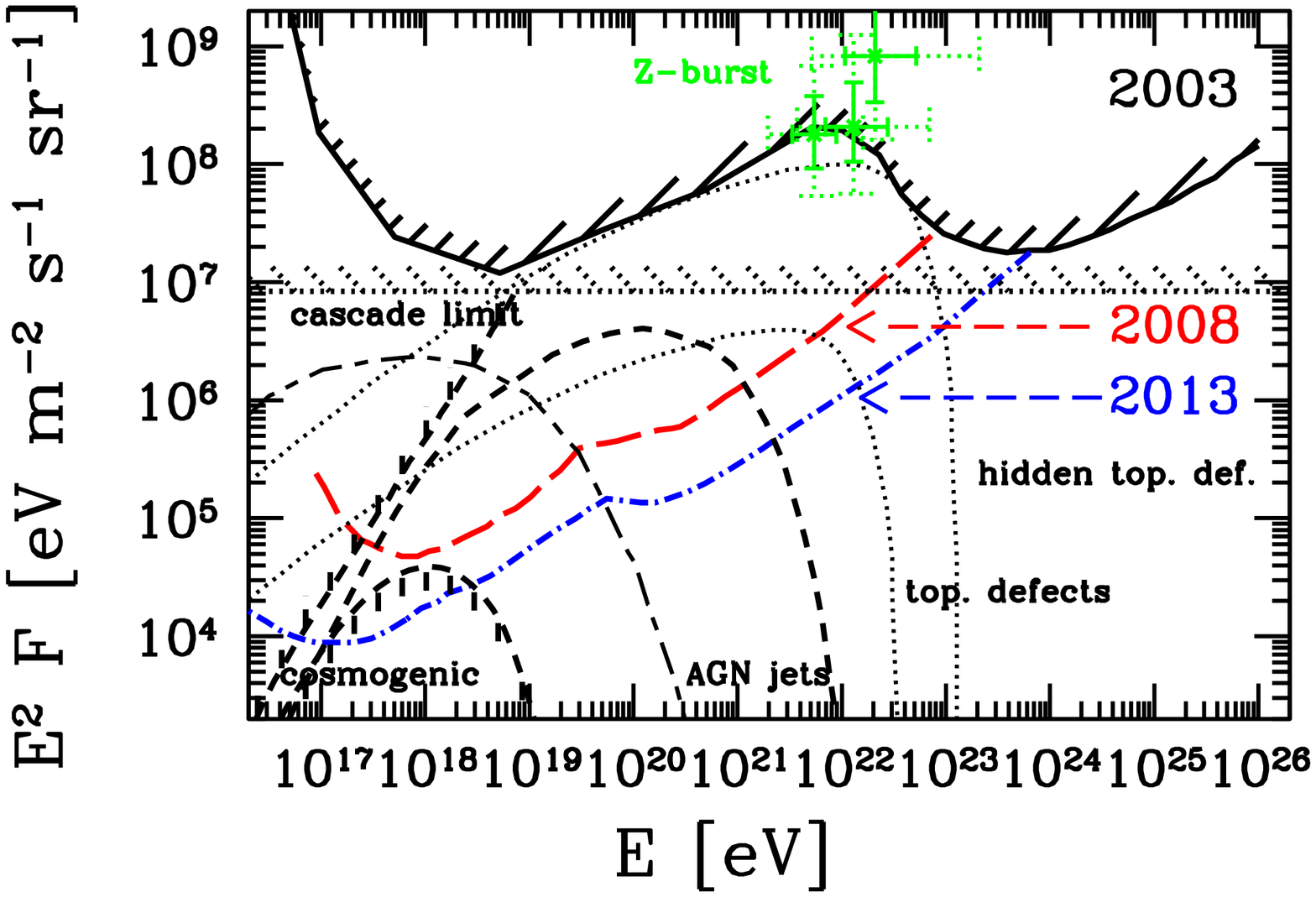}
\vspace{-4ex}
\caption[...]{Current status and next decade prospects for EHEC$\nu$ physics, 
expressed in terms of diffuse neutrino fluxes per flavor, 
$F_{\nu_\alpha}+F_{\bar\nu_\alpha}$, $\alpha =e,\mu,\tau$; 
full mixing among 
the flavors en route to Earth~\cite{Learned:1994wg} is assumed.\\
{\em Top:} Upper limits from  
RICE~\cite{Kravchenko:2003tc}, GLUE~\cite{Gorham:2003da},  
FORTE~\cite{Lehtinen:2003xv}, and Fly's Eye and AGASA~\cite{Baltrusaitis:mt}. 
Also shown are projected sensitivities 
of Auger in $\nu_e$, $\nu_\mu$ modes and in $\nu_\tau$ mode 
(bottom swath)~\cite{Bertou:2001vm}, ANITA~\cite{Gorham:Anita}, EUSO~\cite{Bottai:2003i}, 
and SalSA~\cite{Gorham:private}, corresponding to 
one event per energy decade and indicated duration.\\  
{\em Bottom:} Roadmap for improvement 
in the next decade (2008 and 2013; adapted from Ref.~\cite{Spiering:2003xm}), 
corresponding to one event per energy decade, as well as the current (2003)
observational upper bound (solid-shaded) obtained from Fig.~\ref{roadmap} (top). 
For the year 2008 (long-dashed), we assume 3 yr of Auger data and 
one 15 d ANITA flight.
For 2013 (dashed-dotted), we assume 8/3/4 yr Auger/EUSO/SalSA, 
and 3 ANITA flights. 
The sensitivity will improve if further projects such as Auger North and
OWL~\cite{OWL} are realized, or if the EUSO
flight time is extended. 
Also shown is a wide sample of predictions for EHEC$\nu$ fluxes (discussed in 
\S~\ref{emissivity}).
\label{roadmap}} 
\end{center}
\end{figure}
This resonant energy, when divided by the Z-decay multiplicity of $\sim 40$, 
predicts secondary cosmic ray particles with energies spanning a decade or more above 
the Greisen-Zatsepin-Kuzmin (GZK) energy, $E_{\rm GZK}=4\times 10^{19}$~eV. 
This is the energy beyond which the CMB is absorbing to nucleons, due to 
resonant photopion production~\cite{Greisen:1966jv}. The association of Z-bursts with the mysterious
cosmic rays observed above $E_{\rm GZK}$ is a controversial 
possibility~\cite{Weiler:1999sh,Weiler:Ringberg,Fargion:1999ft,Yoshida:1998it,Fodor:2001qy,Kalashev:2001sh}.
Intriguingly, the neutrino mass window required in such a scenario 
coincides quantitatively with Eq.~(\ref{lim_comb_osc_cosm})~\cite{Fodor:2001qy}. 

\item Recent proposals for progressively larger EHEC$\nu$ detectors, such as 
the Pierre Auger Observatory~\cite{Auger}, IceCube~\cite{IceCube}, ANITA~\cite{ANITA}, 
EUSO~\cite{EUSO}, OWL~\cite{OWL}, and SalSA~\cite{Gorham:2001wr} offer credible hope 
that the collection of an event sample above $10^{21}$~eV may be realized within
this decade~\cite{Spiering:2003xm} (cf. Fig.~\ref{roadmap}). 
Another encouraging sign for the progress in experimental sensitivity is 
that existing EHEC$\nu$ ``observatories'', such as
RICE~\cite{RICE}, GLUE~\cite{GLUE}, and FORTE~\cite{FORTE}, have recently put the first 
sensible upper limits on the 
EHEC$\nu$ flux in the region~(\ref{Eres_lim}) relevant for neutrino 
absorption~\cite{Kravchenko:2003tc,Gorham:2003da,Lehtinen:2003xv}.
We display these limits in Fig.~\ref{roadmap} (top). 

\end{itemize}

The organization of this paper is as follows. In \S~\ref{spectra}, we determine the 
EHEC$\nu$ spectra, with their all-important absorption features, 
to be observed at Earth.
In particular, the depths, widths, and locations 
of the relic neutrino absorption dips are calculated, 
for various proposed sources of a diffuse EHEC$\nu$ flux.  
The experimental prospects to detect the absorption dips 
within the next decade or beyond are discussed in \S~\ref{prospects}.     
Finally, in \S~\ref{conclusions}, we summarize our results and conclude.

\section{\boldmath EHEC$\nu$ spectral dips\label{spectra}}

Given an EHEC$\nu$ source emissivity distribution, one can determine the 
resulting neutrino spectrum to be observed at Earth, by taking into account 
resonant annihilation with the C$\nu$B, and energy losses due to redshift. 
We formulate this problem, in \S~\ref{propagation}, in terms of propagation 
functions~\cite{Yoshida:pt,Fodor:2003ph}.
We calculate these functions 
by means of the procedure outlined in the original papers 
on relic neutrino absorption spectroscopy~\cite{Weiler:1982qy,Weiler:1983xx}, updated to modern cosmological parameters. 
The observable neutrino flux arriving at Earth is then obtained by folding the propagation function
with the EHEC$\nu$ source emissivity distribution. The latter basic input is  
not known in the energy region of interest. 
Therefore, we introduce in \S~\ref{emissivity} 
various parameterizations to model neutrino emission from the most relevant classes of possible sources 
-- astrophysical accelerators (bottom-up ``Zevatrons'')
or exotic massive particles and topological defects (top-down). 
In \S~\ref{case}, we study 
the location, depths, and widths of relic neutrino absorption dips in the 
context of these model classes of sources, and for various neutrino mass scenarios.

\subsection{Neutrino propagation functions\label{propagation}}

Let 
${\mathcal L}_{\nu_\beta}(r,E_i )$ be the EHEC$\nu$ source emissivity distribution, 
i.e.\ the number of neutrinos $\nu_\beta$ of flavor $\beta =e,\mu,\tau$, 
per co-moving volume, per unit of time and per unit of energy as measured at the source,
injected in the C$\nu$B at a ``propagation distance'' $r\equiv c\,t$ from Earth~\footnote{
The convenience of using $c\,t$ as the measure of distance,
with $t$ being the look-back time, is that it is 
easily translated into redshift.
An alternate, but equivalent, derivation of the relation 
between source luminosity and differential flux at Earth,
using not $t$ but rather the comoving distance, is given 
in Ref.~\cite{Weiler:1983xx}.  
Note that, in this original work, the source luminosity is defined
per physical volume, so there is an additional redshift dependence there, traceable to 
${\mathcal L}$(per comoving volume)$=(1+z)^{-3}\,{\mathcal L}$(per physical volume).
}
with an energy $E_i$. The propagation through the C$\nu$B can be    
described~\cite{Yoshida:pt,Fodor:2003ph}
by 
the functions $P_{b|a} (E; E_i,r)$, which are defined as the expected number of 
particles of type $b$ 
above the threshold energy $E$ if one particle of type $a$ 
started at a distance $r$ with energy $E_i$. 
With the help of these propagation functions, 
the differential flux of 
neutrinos of flavor $\alpha$ 
($b=\nu_\alpha$) at Earth, 
i.e. their number $N_{\nu_\alpha}$ arriving at Earth with energy $E$ per units of energy, 
area ($A$), time ($t$) and solid angle ($\Omega$), 
can be expressed as   
\begin{eqnarray}
\label{flux-earth}
\lefteqn{
F_{\nu_\alpha} ( E ) \equiv \frac{{\rm d}^4 N_{\nu_\alpha}}{{\rm d}E\,{\rm d}A\,{\rm d}t\,{\rm d}\Omega}
= }
\\[1.5ex] \nonumber &&
\frac{1}{4\pi}\,
\int\limits_0^\infty {\rm d}E_i \int\limits_0^\infty {\rm d}r 
\,\sum_\beta \,
\frac{-\,\partial P_{\nu_\alpha|\nu_\beta}(E; E_i,r )}{\partial E}
\;{\mathcal L}_{\nu_\beta} (r,E_i)\,,
\end{eqnarray}
where the ``propagation distance'' $r=c\,t$ is related to the redshift $z$ by    
\begin{equation}
\label{dz-dr}
{\rm d}z = (1+z)\,H(z)\,{\rm d}r\,,
\end{equation}
with the evolving Hubble parameter given by
\begin{equation}
\label{H-Omega}
H^2(z) = H_0^2\,\left[ \Omega_{M}\,(1+z)^3 
+\Omega_k\,(1+z)^2 
+ \Omega_{\Lambda}\right]
\,.
\end{equation}
Here, $H_0=h$~100 km/s/Mpc, with 
$h =(0.71\pm 0.07)\times^{1.15}_{0.95}$~\cite{Hagiwara:fs}, 
denotes the present value of the Hubble parameter.  
In Eq.~(\ref{H-Omega}), $\Omega_{M}$, $\Omega_k$, and $\Omega_{\Lambda}$ 
are the present matter, curvature, and vacuum energy densities, 
respectively, in terms of the critical density. 
From the Friedmann equation comes the constraint
that fractional energies must sum to 100\,\%, i.e.\ 
$\Omega_M +\Omega_k +\Omega_\Lambda =1$.
As default values we choose
$\Omega_M = 0.3$, $\Omega_k=0$, and $\Omega_\Lambda = 0.7$. 
These values are collectively known as the ``concordance'' model,
for they fit a wide assortment of 
data~\cite{Spergel:2003cb,Tegmark:2003ud}~\footnote{See, however, Ref.~\cite{Blanchard:2003du}
for a viable alternative to the concordance model. This alternative, which was
obtained from the relaxation of the hypothesis that the primordial fluctuation spectrum, 
as measured in the CMB, can be described by a single power-law, 
has $\Omega_\Lambda =0$, $\Omega_M = 0.88$, 
$\Omega_\nu = 0.12$, and $h=0.46$ as best-fit values.  
Interestingly, the amount of neutrino hot dark  
matter ($\Omega_\nu = 0.12$) needed in this model points to a quasi-degenerate neutrino mass 
spectrum with $m_{\nu_i}\approx 0.8$~eV.}.   

Justified approximations are that 
(i) the only type of energy loss is due to the redshift, 
and 
(ii) the only relevant interaction is absorption on the relic 
neutrino background, dominated by resonant 
Z-production~\footnote{Both approximations, (i) and (ii), 
are very well satisfied in the energy regions of the absorption dips, i.e.\
the energy decade below the resonance energy, on which 
our analysis mainly concentrates. 
At energies well above the Z-resonant values, 
t-channel W and Z exchange becomes a dominant energy loss mechanism.
These t-channel reactions are the focus of Ref.~\cite{Fargion:1999ft}.
}.
With these approximations, 
the differential propagation function is given by 
(see the Appendix for a thorough derivation, 
properly taking into account neutrino mixing effects)
\begin{eqnarray}
\label{neutrino-prop}
\lefteqn{
\frac{-\,\partial P_{\nu_\alpha|\nu_\beta}(E; E_i,z )}{\partial E}
= }
\\[1.5ex] 
\nonumber 
&& 
\delta \left( E - \frac{E_i}{1+z}\right) 
\sum_j \left| U_{\alpha j}\right|^2 P_{\nu_j} (E_i,z) \left| U_{\beta j}\right|^2
,
\end{eqnarray}
where $U_{\alpha j}$ is the leptonic mixing matrix 
and $P_{\nu_j} (E_i,z)$ is the survival probability of a cosmic neutrino $\nu_j$  
injected at a redshift $z$ with energy $E_i=E\, (1+z)$, 
\begin{eqnarray}
\label{surv-prob}
\lefteqn{
P_{\nu_j} (E\, (1+z) ,z) = }
\\[1.5ex] \nonumber && 
\exp
\left[ 
-\int\limits_0^{z} 
\frac{{\rm d}\tilde z}{H(\tilde z)\,(1+\tilde z)}\,
n_{\bar\nu_j}(\tilde z )
\,\sigma^{\rm ann}_{\nu_j\bar\nu_j} (s)\right],
\end{eqnarray}
with 
\beq{s}
s=2\,m_{\nu_j}\,E\,(1+\tilde z)
\,.
\eeq
We remind the careful reader that $z$ is the emission redshift of the cosmic source,
while $\tilde z$ is the redshift at the time of neutrino annihilation,
the latter being integrated from present time back to the emission time.
The exponent in the brackets is (minus) the optical depth (also called the ``opacity'')
for a neutrino emitted at redshift $z$.

The momentum of a massless neutrino today is of order of the CMB energy,
$\sim 3\,T_\gamma \sim 0.7$~meV.
The neutrino momentum at an earlier epoch is then $\sim 0.7\,(1+z)$~meV.
Thus, relic neutrinos will be non-relativistic as long as 
$m_{\nu_j}\gg \langle\,p_{\nu_j}\,\rangle \sim 0.7\,(1+z)\ {\rm meV}$ holds.  
In most of the region in redshift-space which we 
consider, the relic neutrinos are non-relativistic.
This means that their number densities are given by~\cite{Hagiwara:fs}  
\begin{eqnarray}
\nonumber 
n_{\nu_j}(\tilde z ) = n_{\bar\nu_j} (\tilde z ) 
&=& \langle n_\nu\rangle_0 \ (1 + \tilde z )^3 
= \frac{3}{22}\,\langle n_\gamma\rangle_0 \,(1 + \tilde z )^3
\\[1.5ex] \label{relic-dens}
&= & 56\ {\rm cm}^{-3}\,(1 + \tilde z )^3\,,
\end{eqnarray}
with $\langle n_\nu\rangle_0$ and $\langle n_\gamma\rangle_0$ denoting todays (subscript ``0'') 
average number density in relic neutrinos and relic photons, 
respectively~\footnote{Expression~(\ref{relic-dens}) for the relic neutrino density should
be considered as a rather firm prediction. Possible uniform density enhancements
due to lepton asymmetries are negligible in view of the recent, very
stringent bounds on the neutrino degeneracies~\cite{Dolgov:2002ab}.
Moreover, significant ($> 10$) local density enhancements due to gravitational clustering are,  
for the mass range~(\ref{lim_comb_osc_cosm}), only expected in the innermost regions ($< 100$~kpc) 
of very massive ($> 10^{14}\ M_\odot$) -- and hence rare -- clusters~\cite{Singh:2002de}.}. 

The annihilation cross-section $\sigma^{\rm ann}_{\nu_j\bar\nu_j}(s)$, which is 
dominated by the Z-production peak with a rather narrow-width, can be approximated 
by~\cite{Weiler:1982qy}
\begin{equation}
\label{ann-cross}
\sigma^{\rm ann}_{\nu_j\bar\nu_j}(s)
=\langle \sigma^{\rm ann}_{\nu\bar\nu}\rangle \ 
\delta\left( \frac{s}{m_Z^2}-1\right)\,,
\end{equation}
where
\begin{equation}
\langle \sigma^{\rm ann}_{\nu\bar\nu}\rangle = \int \frac{{\rm d}s}{m_Z^2}\;
\sigma^{\rm ann}_{\nu_j\bar\nu_j} 
= 2\pi\sqrt{2}\,G_F 
= 40.4\ {\rm nb}
\end{equation}
is the energy averaged cross-section~\footnote
{If the neutrinos are Dirac particles rather than Majorana particles,
then their transition from relativistic, single-helicity particles to 
non-relativistic unpolarized particles populates the sterile spin-states. 
If such is the case, then the depolarization of the Dirac states halves the 
unpolarized cross-section, which in turn halves the
annihilation rates compared to the rates for Majorana 
neutrinos~\cite{Weiler:1983xx,Weiler:Ringberg}.  
We present results only for the Majorana case, since 
the theoretical models for neutrino-mass generation favor Majorana neutrinos 
(two light spin-states per flavor).
} 
$G_F=1.2\times 10^{-5}$~GeV$^{-2}$ being the Fermi coupling constant~\cite{Hagiwara:fs}.
Therefore, the survival probability~(\ref{surv-prob}), at the injection energy $E_i=E\,(1+z)$,  
can be approximated by~\cite{Weiler:1982qy} 
\begin{widetext}
\begin{eqnarray} 
\label{surv-prob-appr}
P_{\nu_j} (E\,(1+z),z) & \simeq & 
\exp\left[ 
- \, 
\left(\frac{\langle n_\nu\rangle_0\,\langle \sigma^{\rm ann}_{\nu\bar\nu}\rangle}{H_0}\right)\,
\frac{\left(\frac{E_{\nu_j}^{\rm res}}{E}\right)^3}{\left[\Omega_{M}\,
\left(\frac{E_{\nu_j}^{\rm res}}{E}\right)^3 
+ \Omega_k\,\left(\frac{E_{\nu_j}^{\rm res}}{E}\right)^2 
+ \Omega_{\Lambda}\right]^{1/2}}
\right]
,
\\[3.5ex] \nonumber && \hspace{6ex}
{\rm for\ }\ 
\frac{1}{1+z} < \frac{E}{E_{\nu_j}^{\rm res}}<1\,,
\end{eqnarray}
\end{widetext}
within the region of support indicated,
and identically one (no absorption) outside the region of support.
Here, $E_{\nu_j}^{\rm res}$ is the resonant energy in the rest system of the relic neutrinos,
given in Eq.~(\ref{Eres}).

Numerically, the annihilation probability on the C$\nu$B, neglecting cosmological expansion,
is on the few percent level,
\begin{equation}
\label{ann-prob-today}
\frac{\langle n_\nu\rangle_0\,\langle \sigma^{\rm ann}_{\nu\bar\nu}\rangle}{H_0}
=\left( {0.71}/{h}\right) \times 3.0\ \%\ 
\,. 
\end{equation}
Taking cosmological expansion into account, the annihilation probability 
is enhanced by the redshift-dependent
ratio in the exponent appearing in the expression~(\ref{surv-prob-appr}) for the 
survival probability. 
The enhancement is easy to understand:
In the numerator of the ratio, the factor
$(E_{\nu_j}^{\rm res}/E)^3=(1+\tilde z)^3$ accounts for the higher target density of 
the C$\nu$B in physical volume.  The denominator accounts for the 
time or path length available per unit redshift for annihilation.
What is noteworthy here is that, at early times, the 
$\Omega_M$ term dominates, and so the neutrino optical depth 
scales as $1/\sqrt{\Omega_M}$.  Thus, the smaller is $\Omega_M$,
or equivalently, the larger is $\Omega_\Lambda$, the greater is the 
neutrino absorption probability from sources with $1+z\gwig\, (\Omega_\Lambda/\Omega_M)^{1/3}$.
For example, the $\Lambda$CDM universe, with $\Omega_M=0.3$ and $\Omega_\Lambda =0.7$, 
produces absorption dips nearly twice as deep as a pure CDM universe with
$\Omega_M=1$.  In turn, this alleviates the statistics requirement (discussed later)
by a factor of $\sim 3$.
%

\begin{figure}
\begin{center}
\includegraphics*[bbllx=20pt,bblly=221pt,bburx=570pt,bbury=608pt,width=7.7cm]{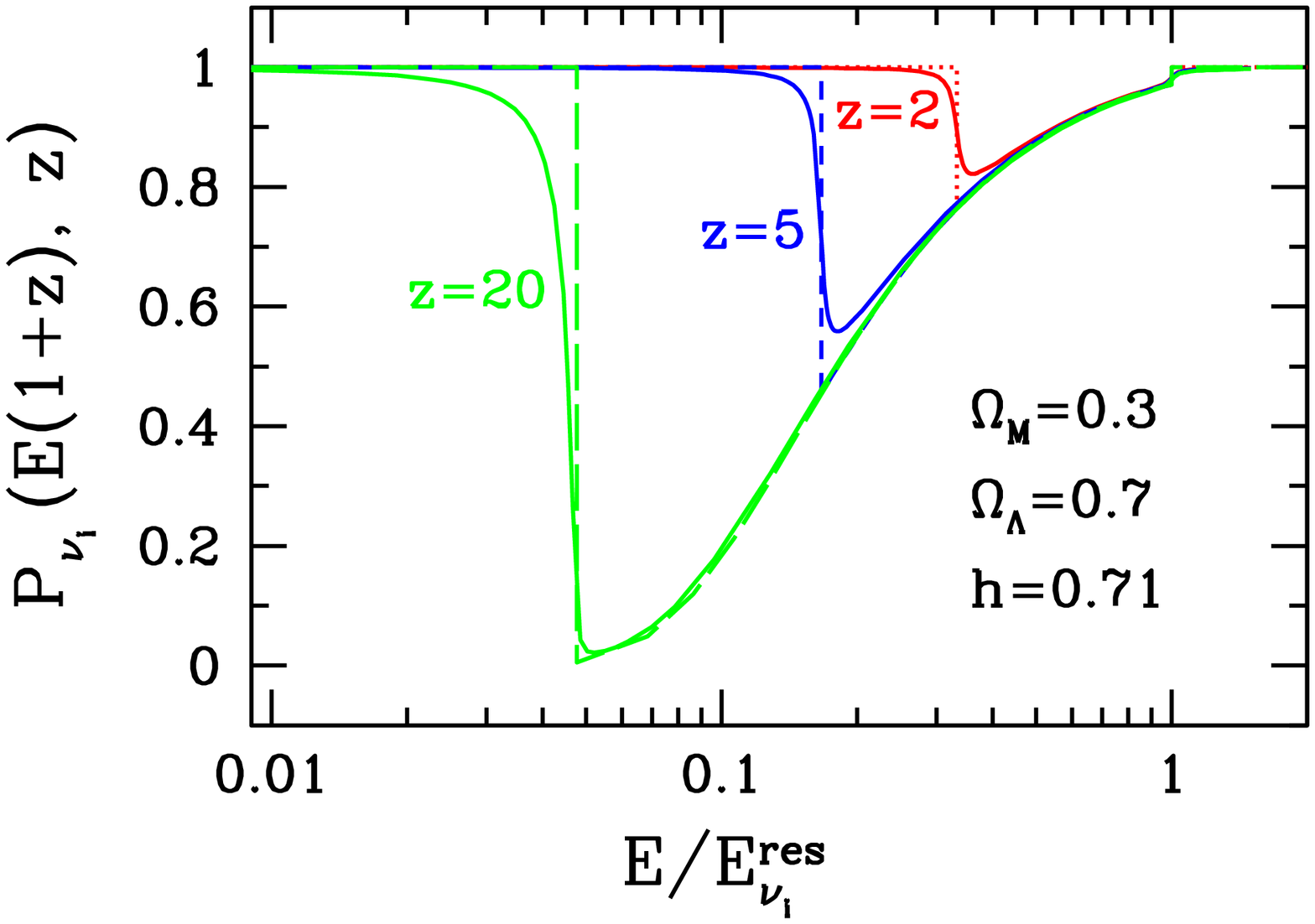}
\includegraphics*[bbllx=20pt,bblly=221pt,bburx=570pt,bbury=608pt,width=7.7cm]{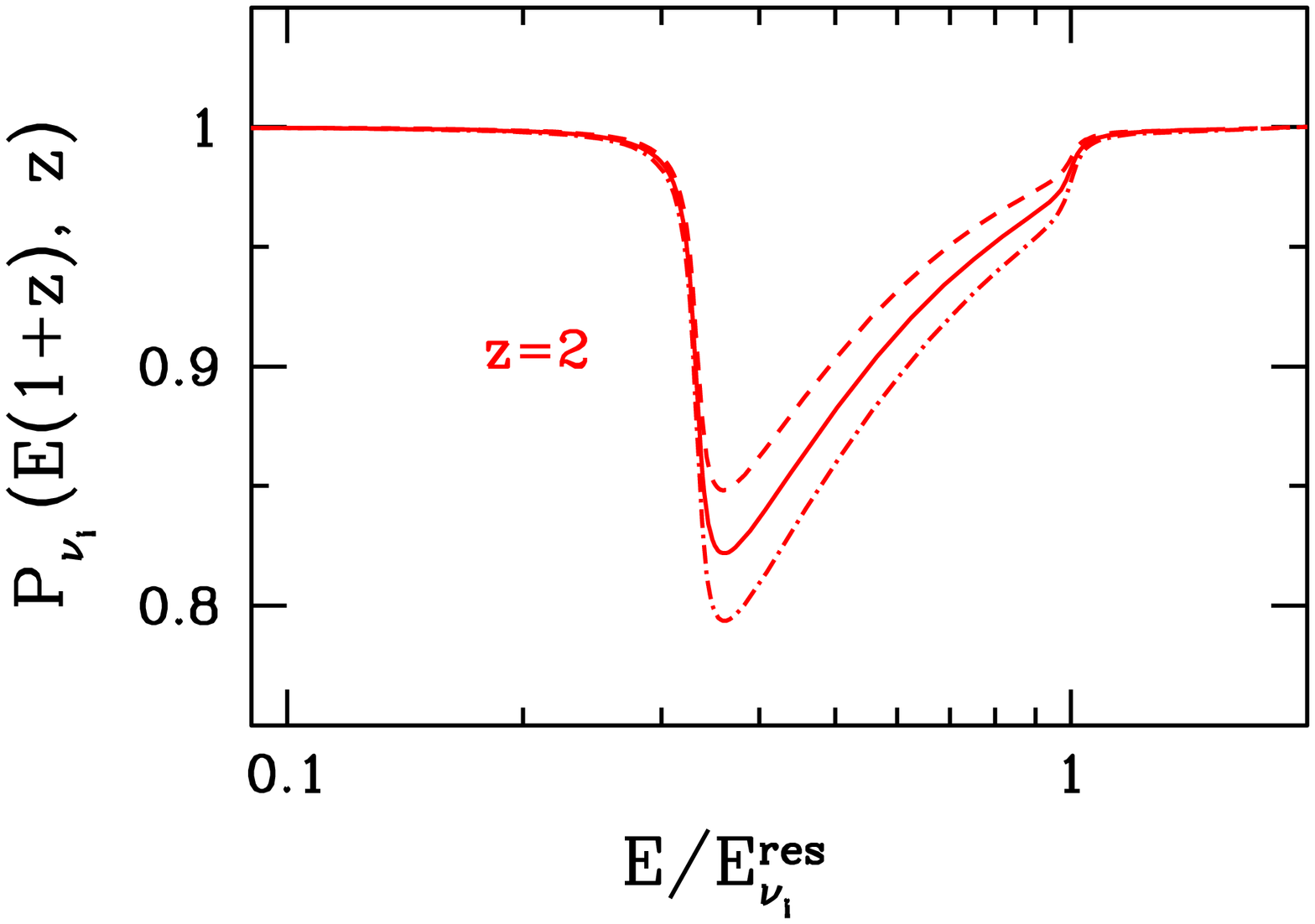}
\caption[...]{The survival probability $P_{\nu_i} (E_i,z)$ of a 
cosmic neutrino $\nu_i$, injected at redshift $z$ with energy $E_i$, 
as a function of the energy at Earth, $E=E_i/(1+z)$, 
in units of the resonance energy $E_{\nu_i}^{\rm res}=m_Z^2/2\,m_{\nu_i}$.\\ 
{\em Top:} The narrow-width approximation~(\ref{surv-prob-appr}), for $z=2$ (dotted), $z=5$ (short-dashed), 
$z=20$ (long-dashed), and standard cosmological parameters 
($\Omega_M=0.3$, $\Omega_\Lambda =0.7$, $h=0.71$),  
compared with 
the complete energy dependence from the annihilation cross-section  of Ref.~\cite{Roulet:1992pz} (solid). \\
{\em Bottom:} The survival probability for $z=2$ and standard cosmological parameters (solid) 
compared with the most extreme variations allowed by up-to-date global fits:
$\Omega_M = 0.20$, $\Omega_\Lambda = 0.78$, $h = 0.81$ (dashed) and 
$\Omega_M = 0.40$, $\Omega_\Lambda = 0.61$, $h = 0.62$ (dashed-dotted)~\cite{Tegmark:2003ud}. 
\label{survival-probability}} 
\end{center}
\end{figure}

In Fig.~\ref{survival-probability}, we display the survival probability~(\ref{surv-prob-appr})  
for the modern concordance cosmological parameters,
i.e.\ $\Omega_M=0.3$, $\Omega_k=0$, $\Omega_\Lambda =0.7$, and $h=0.71$
(top), and their allowed variations (bottom), respectively. 
It seems that variations of $\Omega_M$, $\Omega_\Lambda$, and $h$, 
within their 
uncertainties~\cite{Spergel:2003cb,Tegmark:2003ud}, amount to a $\sim 5$~\% effect. 

As is apparent from Fig.~\ref{survival-probability} (top), 
using the narrow-width approximation~(\ref{ann-cross}) for the annihilation cross-section
rather than taking into account the complete energy dependence of the cross-section 
is justified within the overall $5$~\% error.

\subsection{Neutrino source emissivity distributions\label{emissivity}}

In the previous subsection, we have 
found that, for a given source emissivity distribution  
${\mathcal L}_{\nu_\beta}$ of neutrinos of flavor $\beta$, 
the neutrino flux of flavor $\alpha$ to be observed at Earth is predicted to be 
(cf. Eqs.~(\ref{flux-earth})~-~(\ref{neutrino-prop}))
\begin{eqnarray}
\label{flux-earth-appr}
\lefteqn{F_{\nu_\alpha} ( E ) 
= \frac{1}{4\pi}\,
\int\limits_0^\infty \frac{{\rm d}z}{H(z)} 
\times }
\\[1.5ex] \nonumber && 
\sum_{\beta,\:j}
\left| U_{\alpha j}\right|^2 P_{\nu_j} (E\,(1+z),z) \left| U_{\beta j}\right|^2
\,{\mathcal L}_{\nu_\beta} (z,E\,(1+z))
\,,   
\end{eqnarray} 
where ${\mathcal L}_{\nu_\beta}(z,E(1+z))$ is the number of neutrinos of flavor $\beta$ 
and energy $E_i=E(1+z)$
emitted per co-moving volume, per unit time and unit energy, 
at a redshift ``distance'' $z$ (cf. Eq.~(\ref{dz-dr})). 
In this subsection and the one that follows, 
we evaluate this expression further for some benchmark source emissivities.

We will focus mainly on sources which produce pions (``hadronic'' sources), be they astrophysical Zevatrons
(bottom-up) or non-accelerator (top-down) ones.
From the decay sequence 
$\pi^\pm\rightarrow\mu^\pm +\nu_\mu\rightarrow e^\pm+2\,\nu_\mu+\nu_e$ 
(neither we, nor experiments, will distinguish neutrinos from anti-neutrinos), 
the flavor ratios of the source emissivities are predicted to be 
\begin{equation}
\label{flavor-ratio}
{\mathcal L}_{\nu_e}: {\mathcal L}_{\nu_\mu}: {\mathcal L}_{\nu_\tau}
= 1:2:0
\,.
\end{equation}
In this case, one finds -- exploiting the fact that phenomenologically $\left| U_{e 3}\right|^2\ll 1$ 
and $\left| U_{\mu 3}\right|\simeq \left| U_{\tau 3}\right|$ -- that the fluxes at Earth 
are well approximated by 
(see Appendix and also Ref.~\cite{Learned:1994wg})
\begin{eqnarray}
\label{flux-earth-appr-120}
F_{\nu_\alpha} ( E ) 
&\simeq & \frac{1}{4\pi}\,
\int\limits_0^\infty \frac{{\rm d}z}{H(z)} 
\third\,{\mathcal L}^{\rm tot}_{\nu} (z,E\,(1+z))
\\[1.5ex] \nonumber && \hspace{4ex}\times 
\sum_{j=1}^3
\left| U_{\alpha j}\right|^2 P_{\nu_j} (E\,(1+z),z) 
\,,
\end{eqnarray}
where ${\mathcal L}^{\rm tot}_{\nu}$ is the total, flavor-summed neutrino emissivity at the source. 
In fact, as discussed in the Appendix, Eq.~(\ref{flux-earth-appr-120}) holds 
whenever the source emissivities satisfy 
${\mathcal L}_{\nu_\mu}+{\mathcal L}_{\nu_\tau}=2\,{\mathcal L}_{\nu_e}$.
Thus, it holds also for a ``democratic''
flavor ratio of source emissivities, 
\begin{equation}
\label{flavor-ratio-democratic}
{\mathcal L}_{\nu_e}: {\mathcal L}_{\nu_\mu}: {\mathcal L}_{\nu_\tau}
= 1:1:1
\,,
\end{equation}
as might arise from the decay of topological defects not coupled directly to ordinary 
matter such as, for example, mirror-matter ``necklaces''~\cite{Berezinsky:1999az}. 

\begin{figure}
\begin{center}
\includegraphics*[bbllx=20pt,bblly=159pt,bburx=570pt,bbury=698pt,width=7.cm]{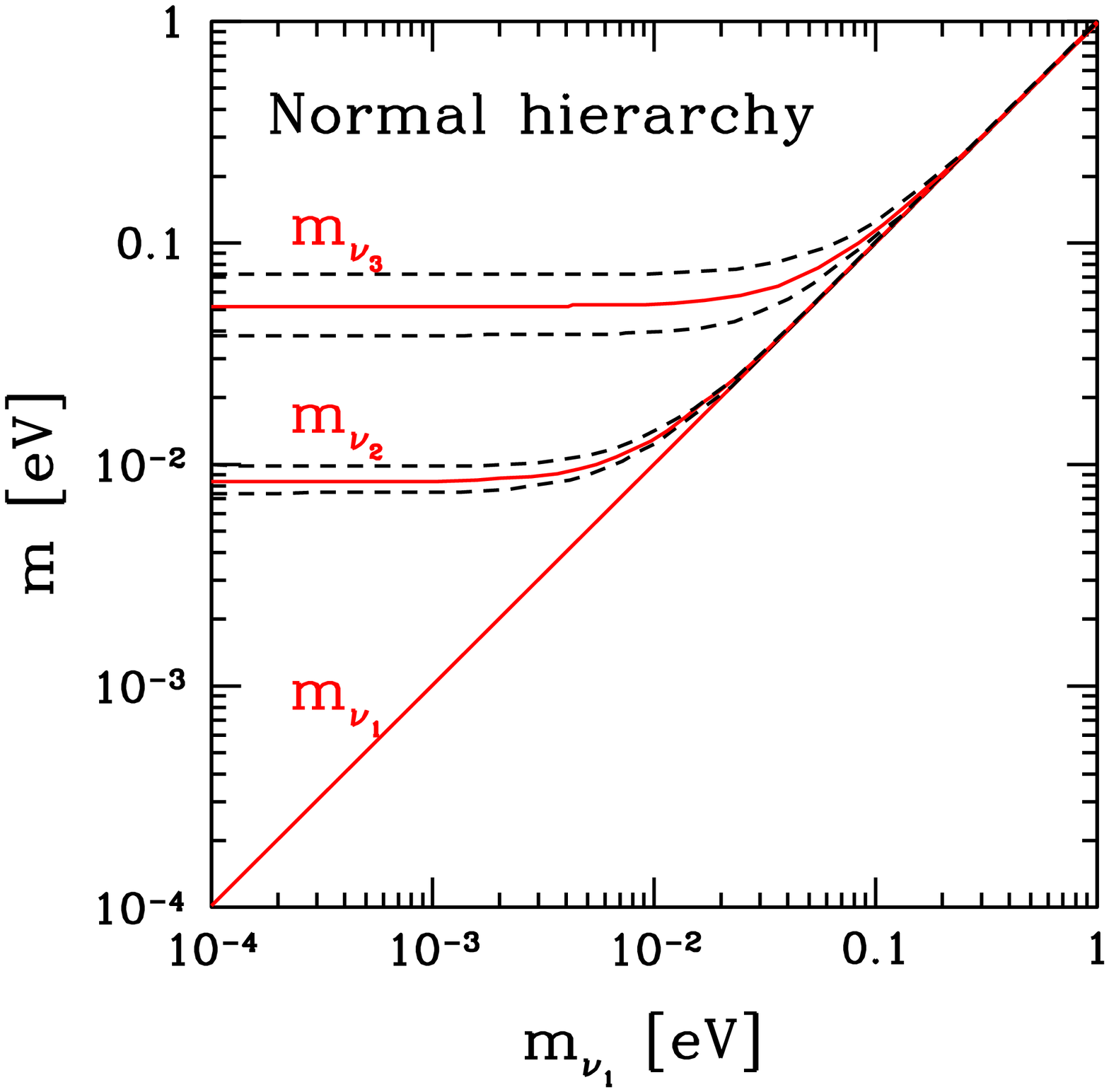}
\includegraphics*[bbllx=20pt,bblly=159pt,bburx=570pt,bbury=698pt,width=7.cm]{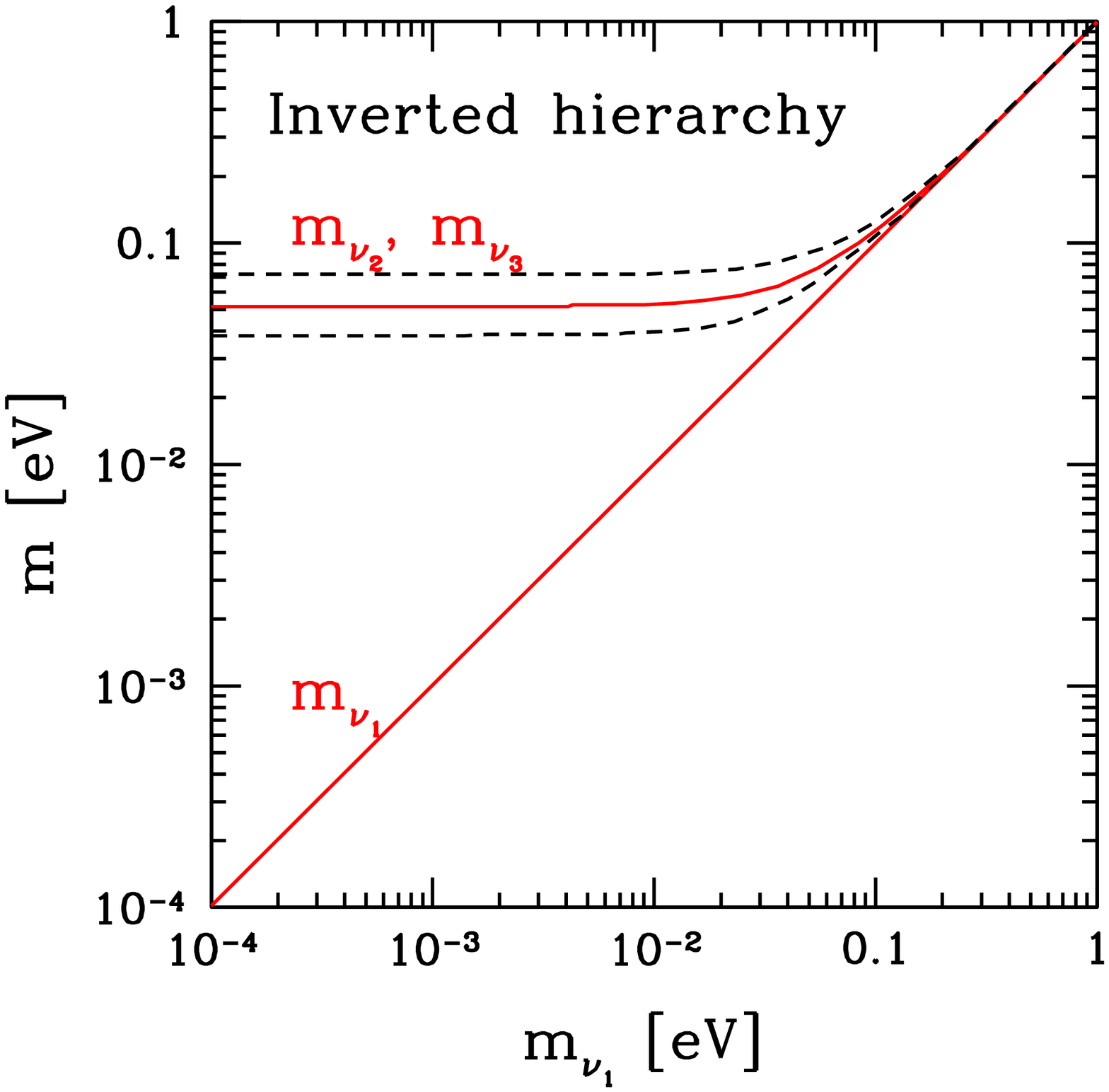}
\caption[...]{Allowed ranges for the neutrino masses as a function of the 
lightest neutrino mass $m_{\nu_1}$, in the normal (top) and 
inverted (bottom) three-neutrino scheme (adapted from Refs.~\cite{Beacom:2002cb,Giunti:2003qt}). 
\label{mass-scen}} 
\end{center}
\end{figure}

It is unlikely that neutrino observatories will be able to distinguish neutrino flavors
at extreme high-energy~\cite{Beacom:2003nh}. 
Phenomenologically then, we are mainly interested in the 
sum over all neutrino flavors $\alpha$. 
With such a sum, unitarity completely removes
the dependence on the leptonic mixing matrix 
from Eq.~(\ref{flux-earth-appr-120}), leaving just
\begin{eqnarray}
\nonumber
\sum_\alpha F_{\nu_\alpha} ( E ) 
&\simeq & \frac{1}{4\pi}\,
\int\limits_0^\infty \frac{{\rm d}z}{H(z)} 
\third\,{\mathcal L}^{\rm tot}_{\nu} (z,E\,(1+z))
\\[1.5ex] \label{flux-earth-appr-120-sum} && \hspace{4ex}\times 
\sum_{j=1}^3 P_{\nu_j} (E\,(1+z),z) 
\,.
\end{eqnarray}

Another simplification of Eq.~(\ref{flux-earth-appr-120})
occurs if the neutrino masses are quasi-degenerate, 
which in fact is realized in Nature if the lightest neutrino has a mass 
$m_{\nu_1}\gg\triangle m_{\rm atm}^2$, say, $m_{\nu_1}\gwig\, 0.1$~eV (cf. Fig.~\ref{mass-scen}):  
\begin{equation}
\label{quasi-deg}
m_{\nu_1}:m_{\nu_2}:m_{\nu_3} \simeq 1:1:1 
\ {\Rightarrow}\  
 P_{\nu_1}\simeq P_{\nu_2}\simeq P_{\nu_3} 
\,.
\end{equation}
With quasi mass-degeneracy, 
Eq.~(\ref{flux-earth-appr-120}) simplifies to 
\begin{eqnarray}
\label{flux-earth-appr-120-deg}
\lefteqn{F_{\nu_\alpha} ( E ) 
\simeq  }
\\[1.5ex] \nonumber &&
\frac{1}{4\pi}\,
\int\limits_0^\infty \frac{{\rm d}z}{H(z)} 
\,P_{\nu_1} (E\,(1+z),z) 
\third\,{\mathcal L}^{\rm tot}_{\nu} (z,E\,(1+z))
\,,
\end{eqnarray} 
for each $\alpha =e,\mu,\tau$.

In the following, we assume that the EHEC$\nu$ sources, which build up the 
diffuse source emissivity, have identical 
injection spectra $J_{\nu_\beta}$. 
Consequently, the $z$ and $E_i$ dependences of the source 
emissivity distribution factorize, 
\begin{equation}
\label{emiss-fact}
{\mathcal L}_{\nu_\beta } (z,E_i) = \eta (z)\,J_{\nu_\beta}(E_i)
\,. 
\end{equation}
The ``activity'' $\eta (z)$ 
is the number of sources per co-moving volume and per unit of time,
at the redshift ``time'' $z$,
while the injection spectra $J_{\nu_\beta}(E_i)$ are
the number of neutrinos $\nu_\beta$ emitted by a single source per unit of energy.
The $z$-dependence of the activity accounts for any  
evolution of the co-moving number density and/or of the common luminosity of the sources. 

We employ several parameterizations of $\eta (z)$, which allow us to study  
a variety of possible EHEC$\nu$ origins 
-- ranging from astrophysical accelerator sources 
such as gamma ray bursts (GRB's) and active galactic nuclei (AGN's) which turned on at $z\sim$~few, 
to non-accelerator sources such as topological defects which have been decaying 
all the way back to $z$ very large.

We start with a parametrization of the activity motivated by astrophysics~\cite{Wick:2003ex},
\begin{equation}
\eta_{\rm SFR} (z) = \eta_0\,\frac{1+a}{(1+z)^{-n_1}+a(1+z)^{n_2}}
\,,
\label{activity-SFR}
\end{equation}
with $\eta_0=\eta(z=0)$ being the activity in the today's epoch.
With the values $a=0.005 (0.0001)$, $n_1=3.3 (4.0)$, and $n_2=3.0 (3.0)$, the parametrization fits  
the star formation rate (SFR) history derived from the blue and UV luminosity
density, in line with the extreme ranges of optical/UV measurements without~\cite{Madau:1997pg} 
(with~\cite{Blain:1999pr}) dust extinction corrections. 
We will refer to these two cases as 
conservative and generous SFR, respectively. 
The star formation rate is believed to map out the earliest structures
sufficiently bound by gravity to undergo fusion.  As such, they may map out the 
history of AGN and GRB evolution, two potential sources of the EHEC rays 
(EHECR's)~\cite{Wick:2003ex}. 

The parametrization~(\ref{activity-SFR}) provides a peak activity at 
\beq{}
1+z_{\rm peak}=\left(\frac{n_1}{a\,n_2}\right)^\frac{1}{n_1+n_2}\,, 
\label{zpeak}
\eeq
which occurs at $z_{\rm peak}=1.4$ 
for the conservative SFR and at $z_{\rm peak}=2.9$ for the generous one.  
For $n_2=0$, $a=0$ it reduces to a simple power-law, 
$\eta(z)=\eta_0\,(1+z)^{n_1}$.
Such a choice, with $n_1\sim 4$ to fit the low-$z$ SFR, is
often used in the literature. 
Without a cutoff in $z$, however, the power-law provides an extreme evolution history,
as the activity increases indefinitely. 

This brings us to a further  
parametrization, namely a simple power-law ansatz,
but with cutoffs $z_{\rm min}$ and $z_{\rm max}$ to exclude the existence of 
nearby and early-time sources, respectively:   
\begin{equation}
\label{activity-pow}
\eta_{\rm pow} (z) =\eta_0\,\left(1+z\right)^n\,
\theta(z-z_{\rm min})\,\theta(z_{\rm max}-z)
\,.
\end{equation}
This ansatz has the advantage of leading to easily trac\-table analytic expressions, while
still reproducing, for $n=4$, the generous SFR case, as long as $z_{\rm max}< z_{\rm peak}$. 
In addition to describing the evolution of GRB's and AGN's up to $z\sim 2$,
the power-law with $n\simeq 1\div 2$ and $z_{\rm max}\gg 1$ 
also characterizes the activity expected from non-accelerator sources.
For example, topological defect sources~\cite{Bhattacharjee:1991zm} are characterized by 
$n=3/2$~\footnote{This is the case for almost all proposed topological defects such as 
ordinary strings, monopolonium, and necklaces. For superconducting cosmic strings it can 
be as large as $n=2$ or bigger.}
and $z_{\rm max}$ arbitrarily large~\footnote{Any realistic injection spectrum 
$J_{\nu_\beta}$ falls off rapidly with very large energy.  
Consequently, the contribution of large $z$ is heavily suppressed
in the relevant $z$-integration~(\ref{flux-earth-appr-120-sum}), 
because the integrand is proportional to $J_{\nu_\beta}(E(1+z))$. Thus, 
the dependence on $z_{\rm max}$ is weak for very large $z_{\rm max}$.}. 
Throughout, we will take $z_{\rm min}=0$ as 
a default value~\footnote{The contribution from small $z$ in Eq.~(\ref{flux-earth-appr-120-sum}) 
is negligible, for any reasonable activity. 
We note that a ``cosmological'' distance of $50$~Mpc corresponds to the small value 
$z= 0.012$, for our default cosmological parameters.}.  

Turning to the source injection spectra $J_{\nu_\beta}(E_i)$,
we only need to specify them in the energy decade around 
the respective resonance energies.
For this energy decade, we again assume a power-law 
behavior~\footnote{For the case of small, i.e.\ non-degenerate or ``hierarchical'', 
neutrino masses 
(cf. Fig.~\ref{mass-scen}), 
the respective resonance energies given in Eq.~(\ref{Eres}) may 
possibly spread over three orders of magnitude. 
In this case, it might be appropriate to consider also 
broken power-law injection spectra. We will not pursue this because, as will be shown below, 
a significant detection of absorption dips in the foreseeable 
future seems to be possible only if neutrino masses are quasi-degenerate.},  
\begin{equation}
\label{inj-spect-pow}
J_{\nu_\beta} (E_i) = j_{\nu_\beta}\,E^{-\alpha}_i\,
\theta(E_i-E_{\rm min})\,\theta(E_{\rm max}-E_i)
\,. 
\end{equation}
As with the activity power-law,
this spectral power-law parametrization facilitates the analytic study 
of different scenarios for the origin of EHEC$\nu$'s. 

In most of the source models, the neutrinos originate from 
pion decays, the latter either being produced in inelastic $pp$ or 
$p\gamma$ interactions (astrophysical sources), or, 
alternatively, arising in the fragmentation of QCD jets 
in the decays of superheavy particles (top-down sources). Although the resulting 
neutrino spectra $J_{\nu_\beta}$ can be fairly well calculated for given injection 
spectra of protons or pions, the details are tedious and not worth  the effort 
for our present purposes,  
and so we retain the simple parametrization~(\ref{inj-spect-pow}) instead. 
The power-law should mimic the various neutrino injection spectra reasonably well,
with the spectral index in the range $\alpha\simeq 1\div 2$. 
Throughout, we will
take $E_{\rm min}=0$ as a default value.

For astrophysical accelerators, one expects that $E_{\rm max}$ is 
a fraction ($\sim 5$~\%) of the maximum proton energy, $E_{p\,{\rm max}}$.
In the case of shock acceleration, $E_{p\,{\rm max}}$ is determined
by the requirement that the gyromagnetic radius of the protons in the ambient 
magnetic field $B$ be less than
the spatial extension $R$ of the accelerating source. The result is
\begin{equation}
\label{fermi-acc}  
E^{\rm shock}_{p\ {\rm max}}\simeq 10^{21}\ {\rm eV}\ (R/{\rm kpc})\,(B/{\rm mG})\,.
\end{equation}
Even higher energies are possible in proposed non-shock mechanisms,
such as unipolar induction, acceleration in strong electromagnetic waves 
in plasmas (wakefields)~\cite{Chen:2002nd}, or by magnetic recombination in the vicinity 
of massive black holes~\cite{Li:1998yg}.

For top-down scenarios, $E_{\rm max}$ can be much larger, basically bounded only by the 
huge mass of the decaying particle or defect.
These huge masses are thought to reflect the energy-scale of an 
underlying phase transition.
Popular examples include $M_{\rm GUT}\sim 10^{16}$~GeV from grand-unification, 
and $M_{\rm wimpzilla}\sim 10^{11\div 13}$~GeV from the end of inflation.
Neutrinos from the decay of these super-massive objects carry an 
energy about one order of magnitude less than the mass.

In general, there may be several classes of sources -- such as GRB's, AGN's, EHEC protons 
scattering inelastically off the CMB (``cosmogenic'' neutrinos), and  
topological defects --  
which build up the total emissivity distribution. 
Predictions from a variety of such source classes are shown in Fig.~\ref{roadmap} (bottom):
from jets of AGN's~\cite{Mannheim:mm}, from ordinary topological defects ($M_X=10^{14}$~GeV)~\cite{Kalashev:2002kx},  
from hidden-sector topological defects ($M_X=4\times 10^{14}$~GeV)~\cite{Berezinsky:2003iv}, and  
from the Z-burst scenario normalized to fit the highest energy cosmic ray anomaly~\cite{Fodor:2001qy}
with different assumptions about the universal radio background. 
Also shown are the upper and lower bounds~\cite{Fodor:2003ph} (short-dashed-shaded) 
and one example~\cite{Kalashev:2002kx} 
(short-dashed) of the cosmogenic neutrino flux.  
Finally, the cascade limit~\cite{Berezinsky:1975} from Ref.~\cite{Mannheim:1998wp}
(dotted-shaded) on transparent  neutrino sources (discussed later) is shown.
It applies to all scenarios where neutrinos originate from pion decays 
or even from electroweak jets~\cite{Berezinsky:2002hq}. These neutrinos are accompanied by photons 
and electrons which cascade down in energy during their propagation through the universe. 
The cascade limit arises from the requirement that 
the associated diffuse gamma-ray fluxes should not exceed measurements~\footnote{
\label{casc}The cascade limit from Ref.~\cite{Mannheim:1998wp} exploits the 
measurement of the diffuse $\gamma$ ray background from 30 MeV to 100 GeV by EGRET~\cite{Sreekumar:1997un}.
A lower extragalactic contribution to the $\gamma$ ray 
background than that inferred in Ref.~\cite{Sreekumar:1997un}, by roughly a factor of two, 
has been proposed recently~\cite{Strong:2003ex}. 
The cascade limit may therefore be stronger by a corresponding factor.
Also, in the next few years the GLAST experiment~\cite{GLAST}, and eventually its successors,
e.g.\ Constellation-X~\cite{Constellation-X}, may resolve some of the diffuse flux, thereby lowering 
the cascade limit even further.}.

It is straightforward to generalize Eq.~(\ref{emiss-fact}) to a sum of source classes.
In reality, it is more prudent to wish for even one class of sources at the 
energies of interest here, $10^{22\div 23}$~eV,
and so we continue to work within the single source-class hypothesis.

Another possibility, in principle, is to collect events from a 
small number of point sources, possibly just one.
This presents the advantage that the spectra are not smeared by the
integration over the redshift distribution.
Since neutrinos are not deflected in flight, point-source selection is possible.
However, we will find that the event numbers required for statistical 
significance is on the order of 100 or more.  This large number 
presents a luminosity challenge for a small number of 
very bright sources~\cite{Weiler:1983xx}.
In the end, it does not matter much for our purposes whether the EEC$\nu$ flux is 
diffuse of granular, since the isotropy of the relic background ensures nearly the same
absorption shapes in the spectrum of either source-type.

\subsection{Case studies\label{case}} 

Let us start our analyses with the quasi-degenerate neutrino mass-spectrum case,
whose implications are summarized in Eq.~(\ref{quasi-deg}). 
In view of the expectation that EHEC$\nu$ neutrino fluxes are rapidly falling with
energy (cf. Fig.~\ref{roadmap}), this case has the best prospects of observability, since it corresponds to the
largest 
neutrino masses (cf. Fig.~\ref{mass-scen}), and it brings all three resonance energies to 
a common lowest value~(\ref{Eres}).     
As discussed above (cf.\ Eq.~(\ref{flux-earth-appr-120-deg})), we expect in this case an identical
absorption dip in the neutrino flux of each flavor to be observed at Earth, if the 
flavor ratios 
at the sources were 
hadron-like~(\ref{flavor-ratio}) or 
democratic~(\ref{flavor-ratio-democratic}). 

\begin{figure}
\begin{center}
\includegraphics*[bbllx=20pt,bblly=221pt,bburx=570pt,bbury=626pt,width=8.65cm]{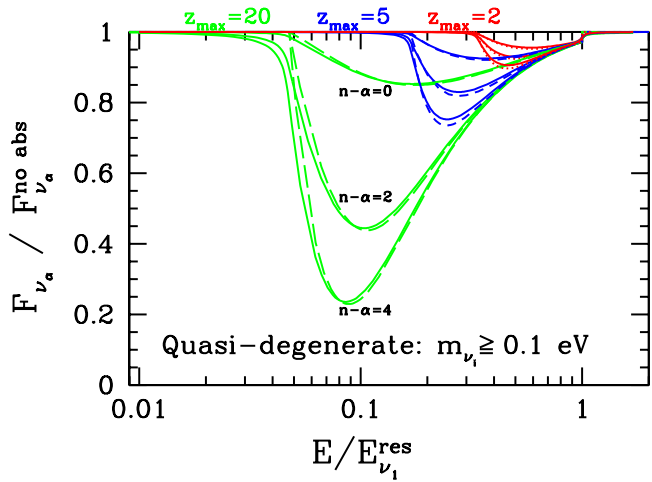}
\caption[...]{Predicted flux~(\ref{flux-earth-appr-120-deg}) 
of neutrinos $\nu_\alpha$ of flavor $\alpha =e,\mu,\tau$ at Earth, 
for a source emissivity characterized by a power-law activity~(\ref{activity-pow}) and 
a power-law injection spectrum~(\ref{inj-spect-pow}), 
for the case of quasi-degenerate neutrino
masses~(\ref{quasi-deg}), normalized to the predicted flux for no absorption.
$E/E_\nu^{\rm res}$ scales as the degenerate mass $m_\nu$.
Three values of $z_{\rm max}$ are shown: 2 (dotted), 5 (short-dashed), and 20 (long-dashed).
For each choice of $z_{\rm max}$, three choices of 
$n-\alpha$ are shown: 0 (upper), 2 (middle), and 4 (lower).  
The corresponding solid lines show the same quantity evaluated with the complete  
energy dependence of the annihilation 
cross-section from Ref.~\cite{Roulet:1992pz} arising from the finite Z-width, instead of 
exploiting the zero-width approximation~(\ref{ann-cross}). 
For all curves, $E_{\rm max}>E_{\nu_1}^{\rm res}\,(1+z_{\rm max})$ is assumed.
\label{absorpt-pow-deg}} 
\end{center}
\end{figure}

In Fig.~\ref{absorpt-pow-deg}, we show the predicted flux $F_{\nu_\alpha}$ 
for a 
hadronic or flavor democratic source emissivity characterized by a power-law activity~(\ref{activity-pow}) 
with redshift evolution index $n$, and power-law injection spectrum~(\ref{inj-spect-pow}) 
with index $\alpha$.  This flux is normalized to the predicted flux for no 
absorption, $F_{\nu_\alpha}^{\rm no\ abs}$    
(obtained by replacing the survival probability in Eq.~(\ref{flux-earth-appr-120-deg}) by unity). 
A nice feature following from the two power-laws, source activity and source emissivity, 
is that the normalized spectrum depends on $n$ and $\alpha$ only through the combination $n-\alpha$;
the source evolution and the energy fall-off compensate each other in a simple way. 
Three particular $n-\alpha$ combinations are shown in Fig.~\ref{absorpt-pow-deg} for
each fixed $z_{\rm max}$: 
$n-\alpha = 0$ (upper curves), $n-\alpha = 2$ (middle curves), and $n-\alpha = 4$ (lower curves).

Figure~\ref{absorpt-pow-deg} illustrates some general features.
Viewed as a function of decreasing energy, 
the absorption dip starts abruptly at the resonance energy $E_{\nu_i}^{\rm res}$. 
This initial depth is determined by today's annihilation probability, $\approx 3$~\%
according to (\ref{ann-prob-today}), 
and reflects the absorptions occurring in nearby ($z\sim 0$) space.
The dip extends in energy down to $E_{\nu_i}^{\rm res}/(1+z_{\rm max})$,
with this minimum energy being the redshifted value of the resonant energy for annihilations
occurring at cosmic distance ($z\sim z_{\rm max}$).
The overall depth of the dip increases dramatically with $n-\alpha$,
thereby showing a strong preference for source evolution and/or flat energy-spectra.
With increasing $z\sim z_{\rm max}=$~2, 5, 20, the absorption depths are roughly 
5, 8, and 15~\% for $n-\alpha = 0$; 7, 18, and 55~\% for $n-\alpha = 2$; and 
10, 27, and 77~\% for $n-\alpha = 4$, respectively.
We observe in Fig.~\ref{absorpt-pow-deg}, 
as we did in Fig.~\ref{survival-probability} (top),
that the replacement of the finite Z-width with
the $\delta$-function approximation~(\ref{ann-cross}) 
is well justified for our purposes. 
From now on, we will exploit this simplification.

\begin{figure}
\begin{center}
\includegraphics*[bbllx=20pt,bblly=221pt,bburx=570pt,bbury=608pt,width=7.2cm]{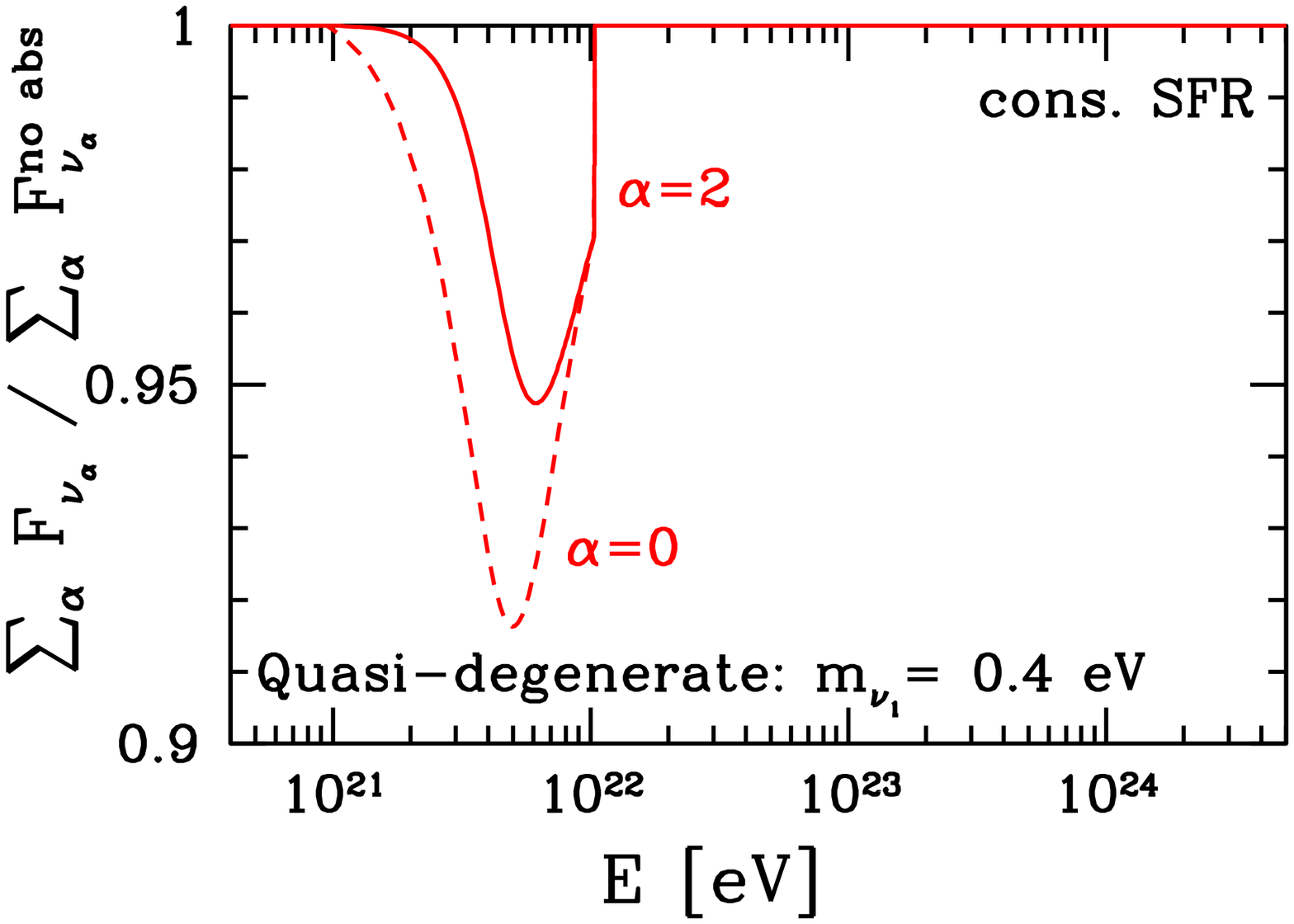}
\includegraphics*[bbllx=20pt,bblly=221pt,bburx=570pt,bbury=608pt,width=7.2cm]{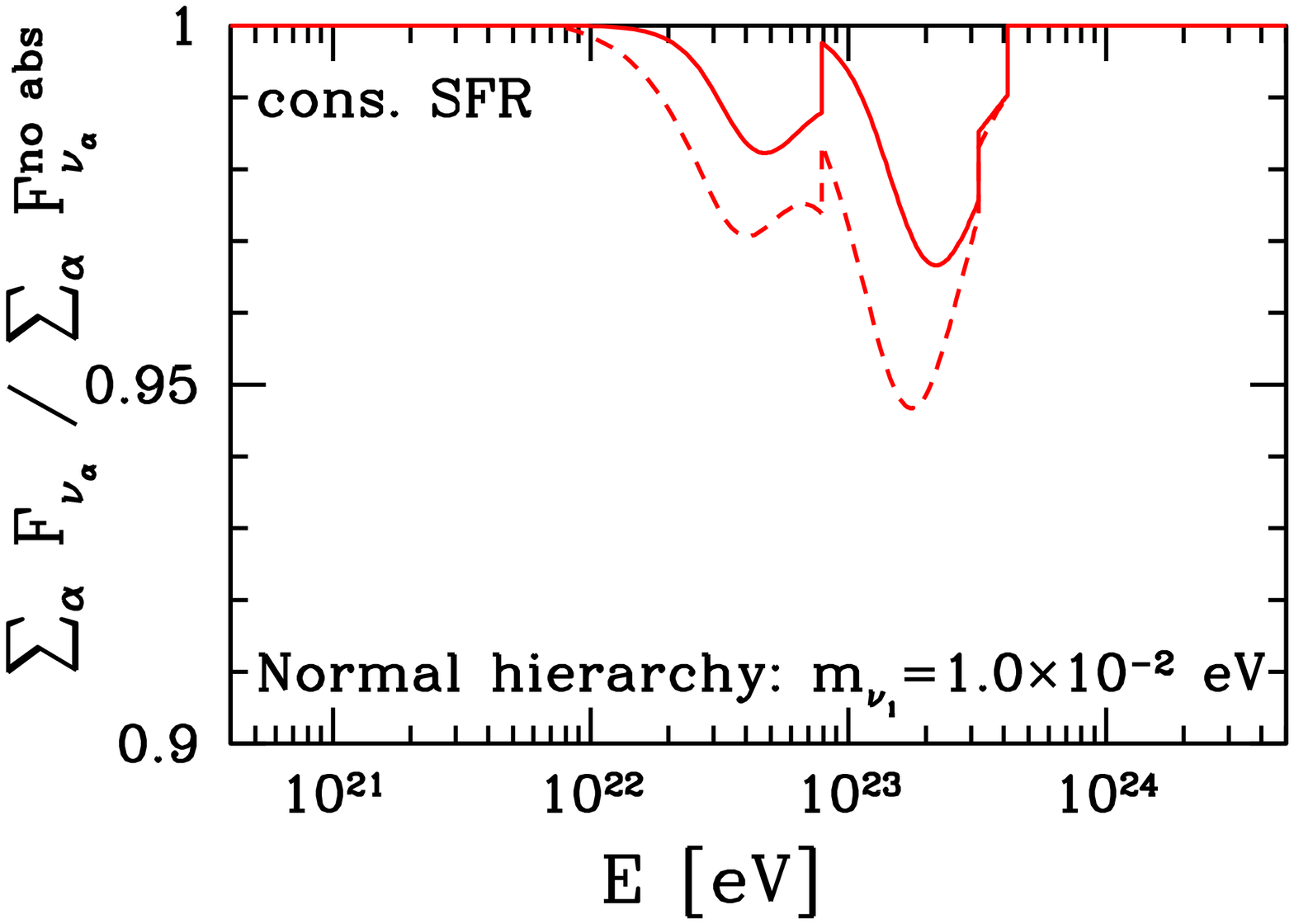}
\includegraphics*[bbllx=20pt,bblly=221pt,bburx=570pt,bbury=608pt,width=7.2cm]{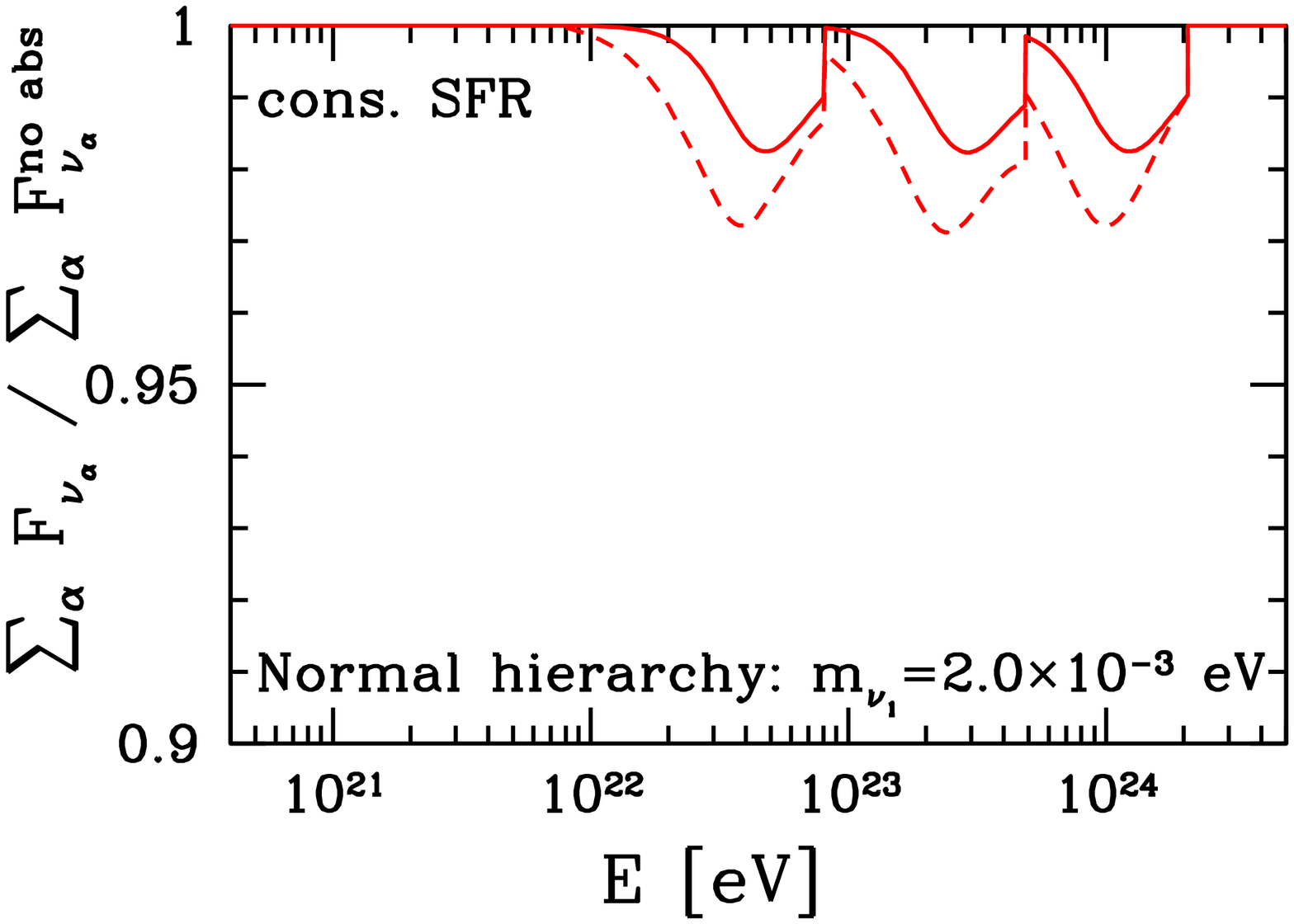}
\includegraphics*[bbllx=20pt,bblly=221pt,bburx=570pt,bbury=608pt,width=7.2cm]{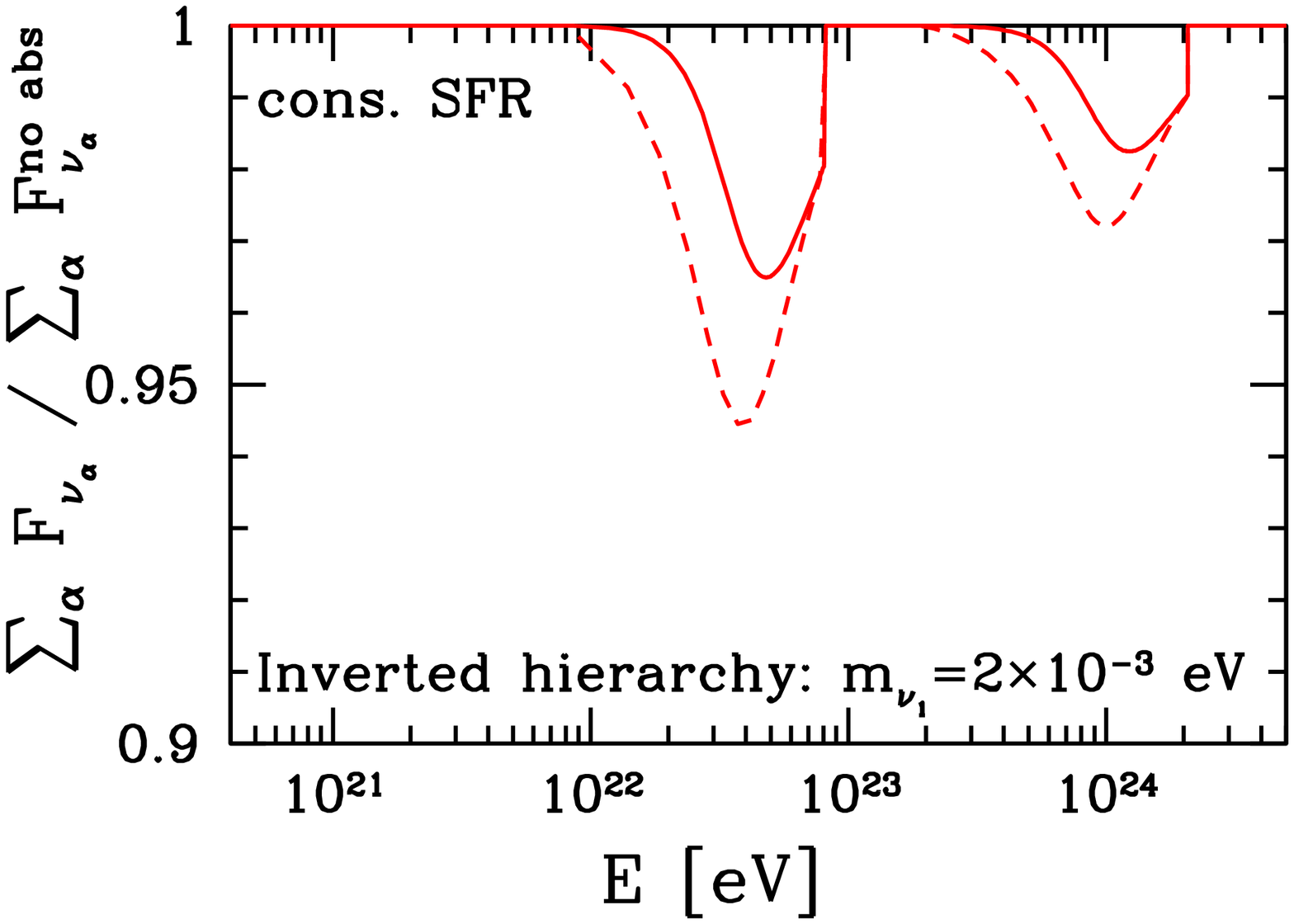}
\caption[...]{Predicted flux 
of neutrinos summed over flavors at Earth~(\ref{flux-earth-appr-120-sum}), 
normalized to the predicted flux for no absorption, 
for a conservative 
SFR activity~(\ref{activity-SFR}) 
to $z_{\rm max} =10$,
injection-spectrum indices $\alpha = 2$ (solid) and $\alpha = 0$ (dashed)~(\ref{inj-spect-pow}), 
and neutrino spectra which are quasi-degenerate (top), normal-hierarchical (2nd and 3rd panels), 
and inverted-hierarchical (bottom panel).
For all curves it is assumed that $E_{\rm max}>E_{\nu_1}^{\rm res}\,(1+z_{\rm max})$. 
\label{absorpt-consSFR-norm}} 
\end{center}
\end{figure}

\begin{figure}
\begin{center}
\includegraphics*[bbllx=20pt,bblly=221pt,bburx=570pt,bbury=608pt,width=7.2cm]{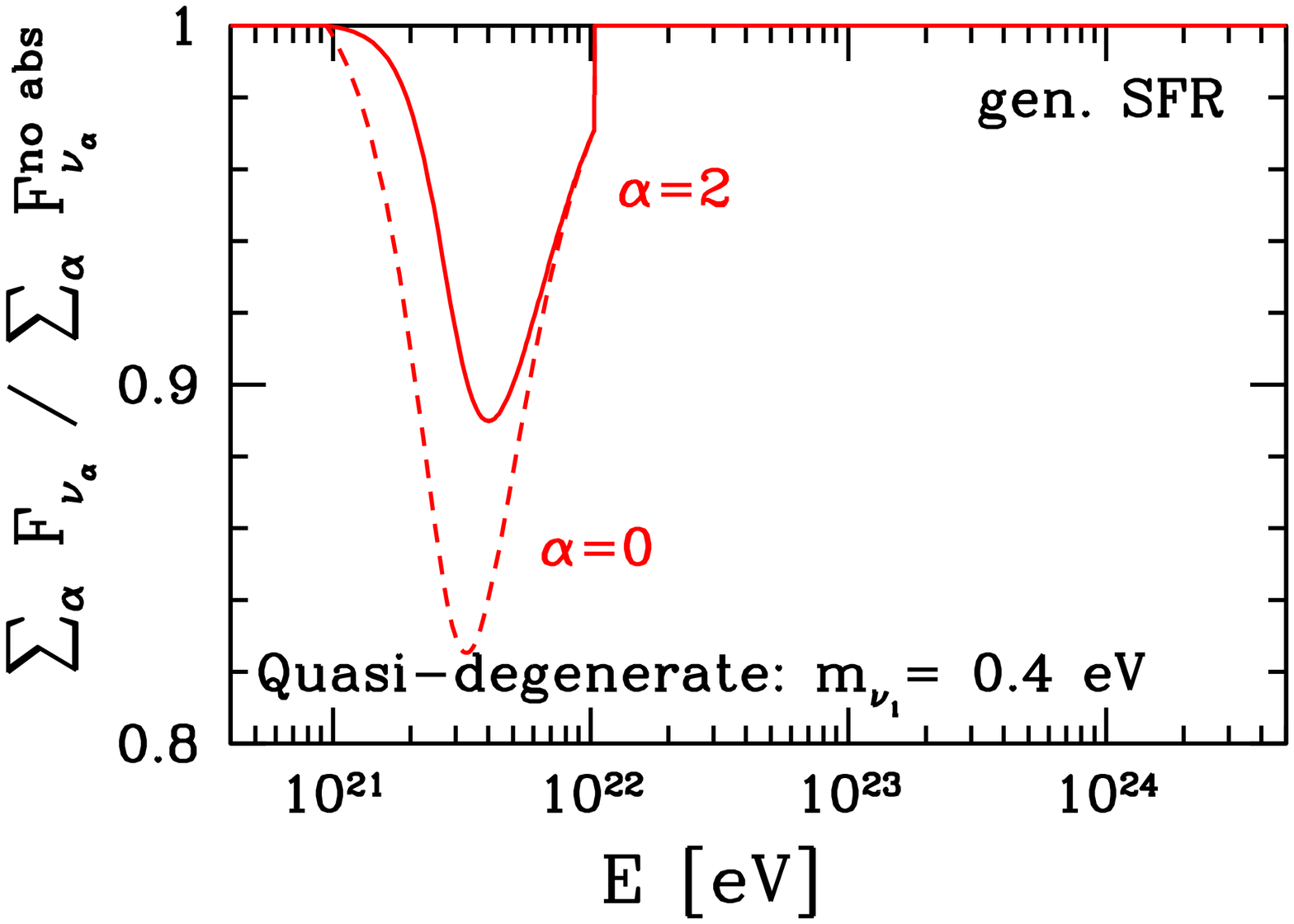}
\includegraphics*[bbllx=20pt,bblly=221pt,bburx=570pt,bbury=608pt,width=7.2cm]{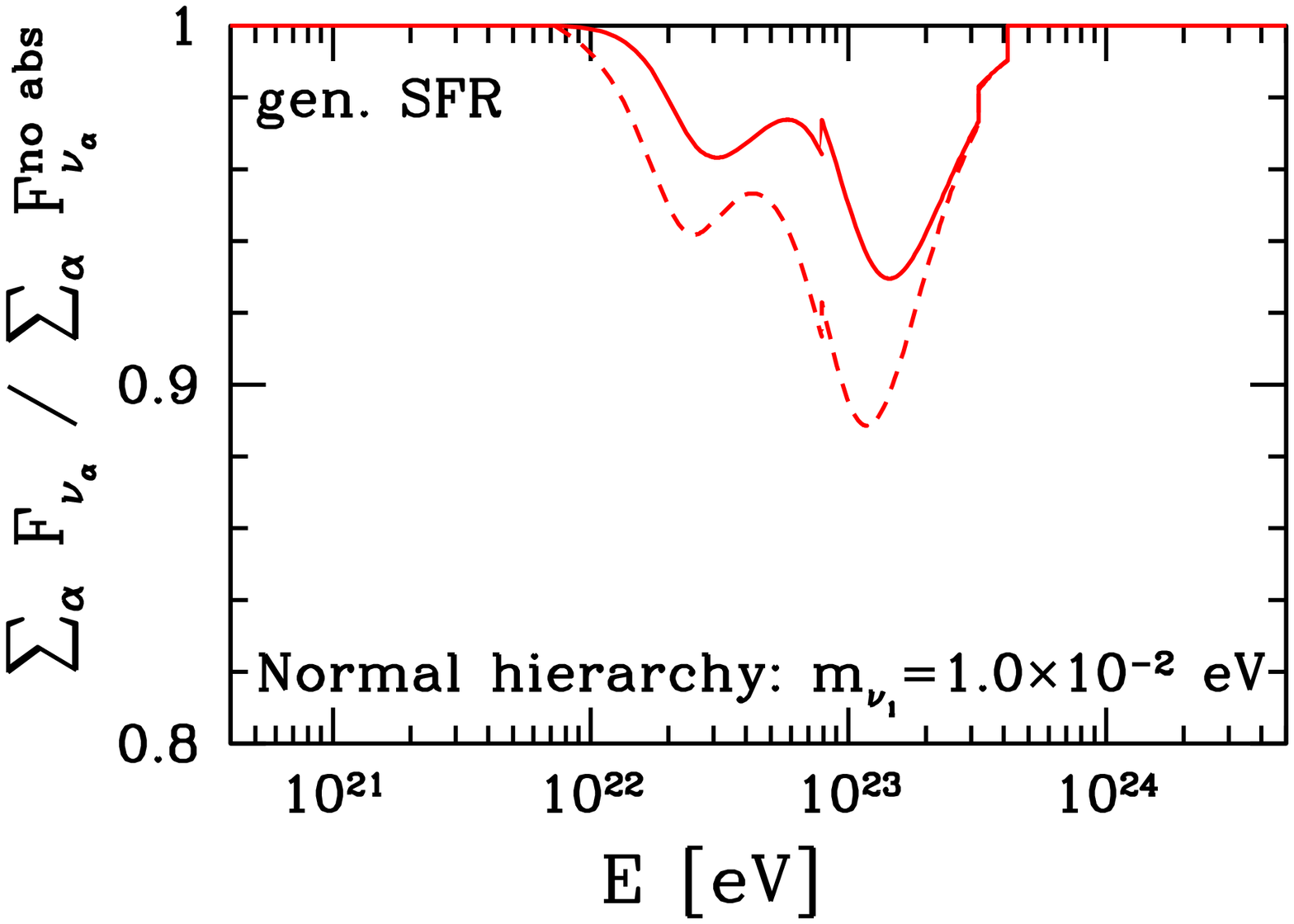}
\includegraphics*[bbllx=20pt,bblly=221pt,bburx=570pt,bbury=608pt,width=7.2cm]{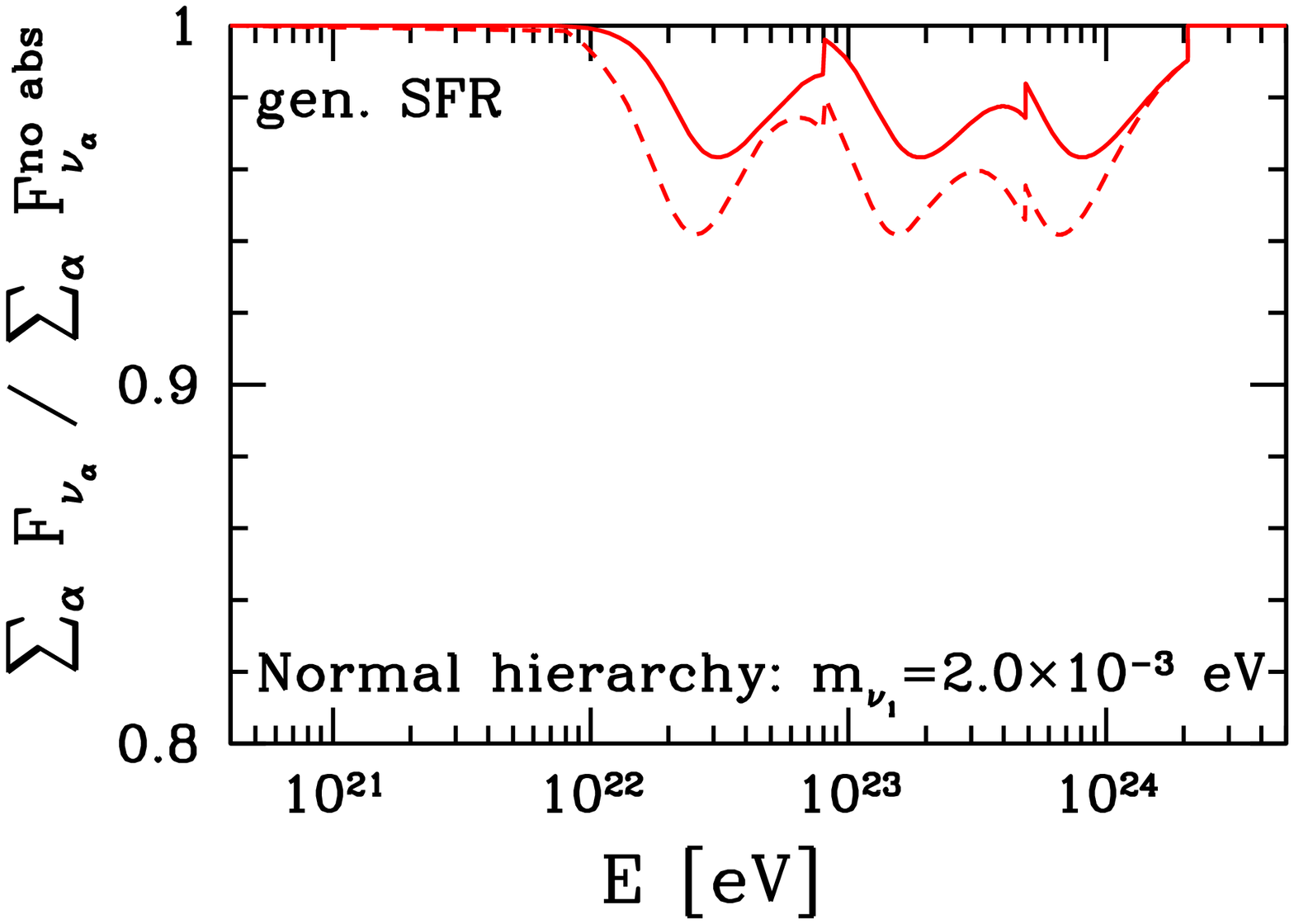}
\includegraphics*[bbllx=20pt,bblly=221pt,bburx=570pt,bbury=608pt,width=7.2cm]{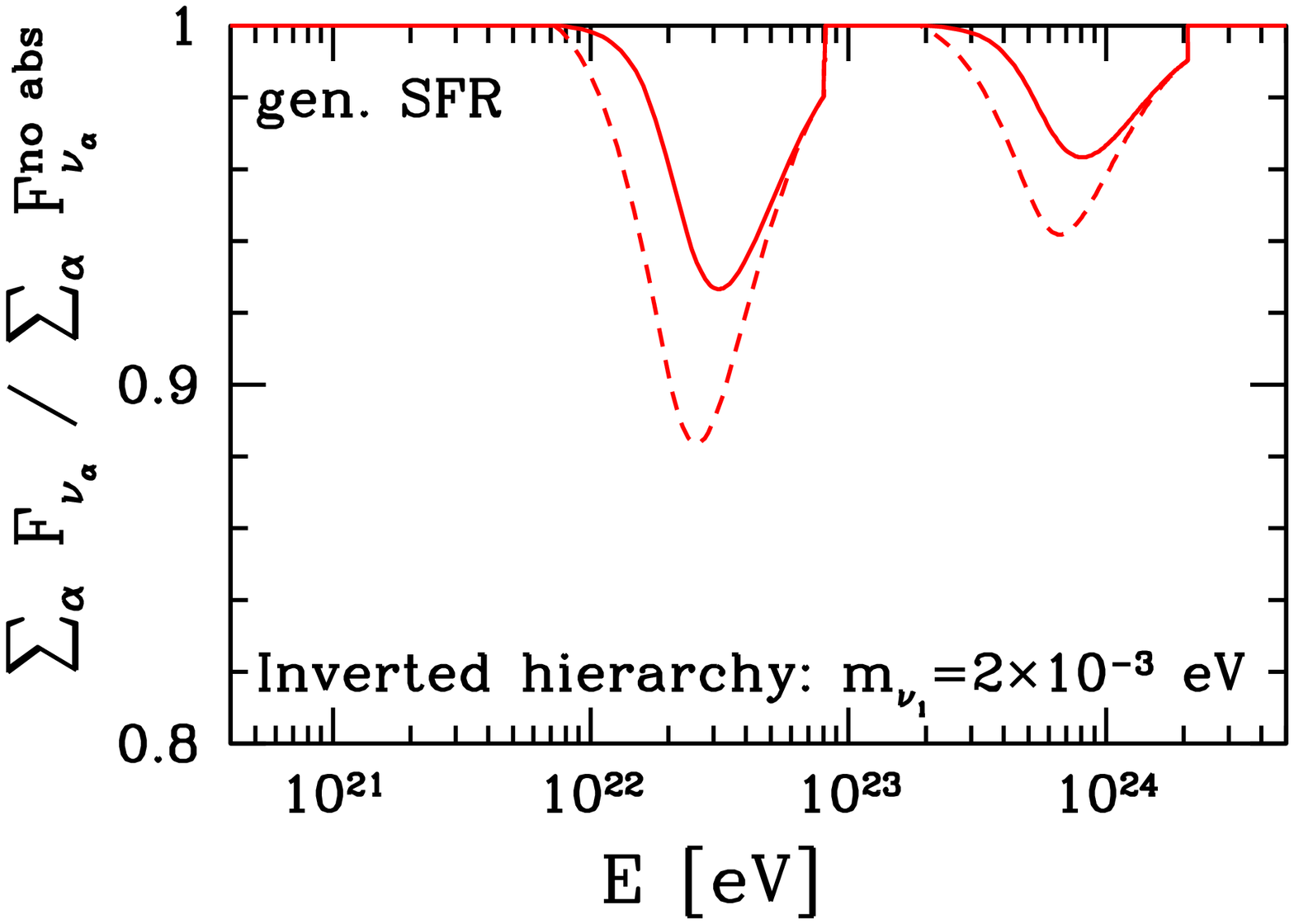}
\caption[...]{Same as Fig.~\ref{absorpt-consSFR-norm}, but with the generous SFR activity~(\ref{activity-SFR}).
\label{absorpt-genSFR-norm}} 
\end{center}
\end{figure}

\begin{figure}
\begin{center}
\includegraphics*[bbllx=20pt,bblly=221pt,bburx=570pt,bbury=608pt,width=7.2cm]{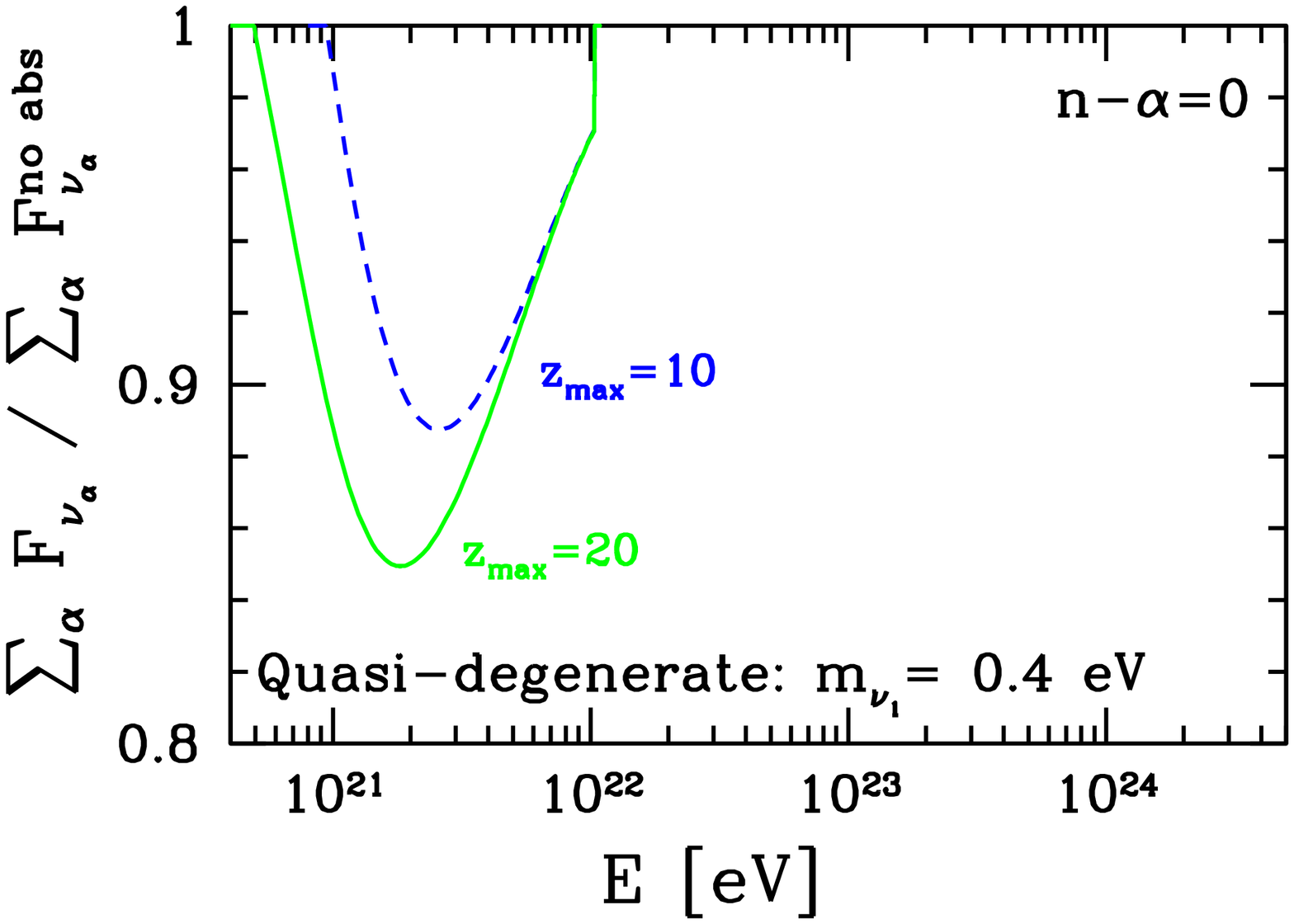}
\includegraphics*[bbllx=20pt,bblly=221pt,bburx=570pt,bbury=608pt,width=7.2cm]{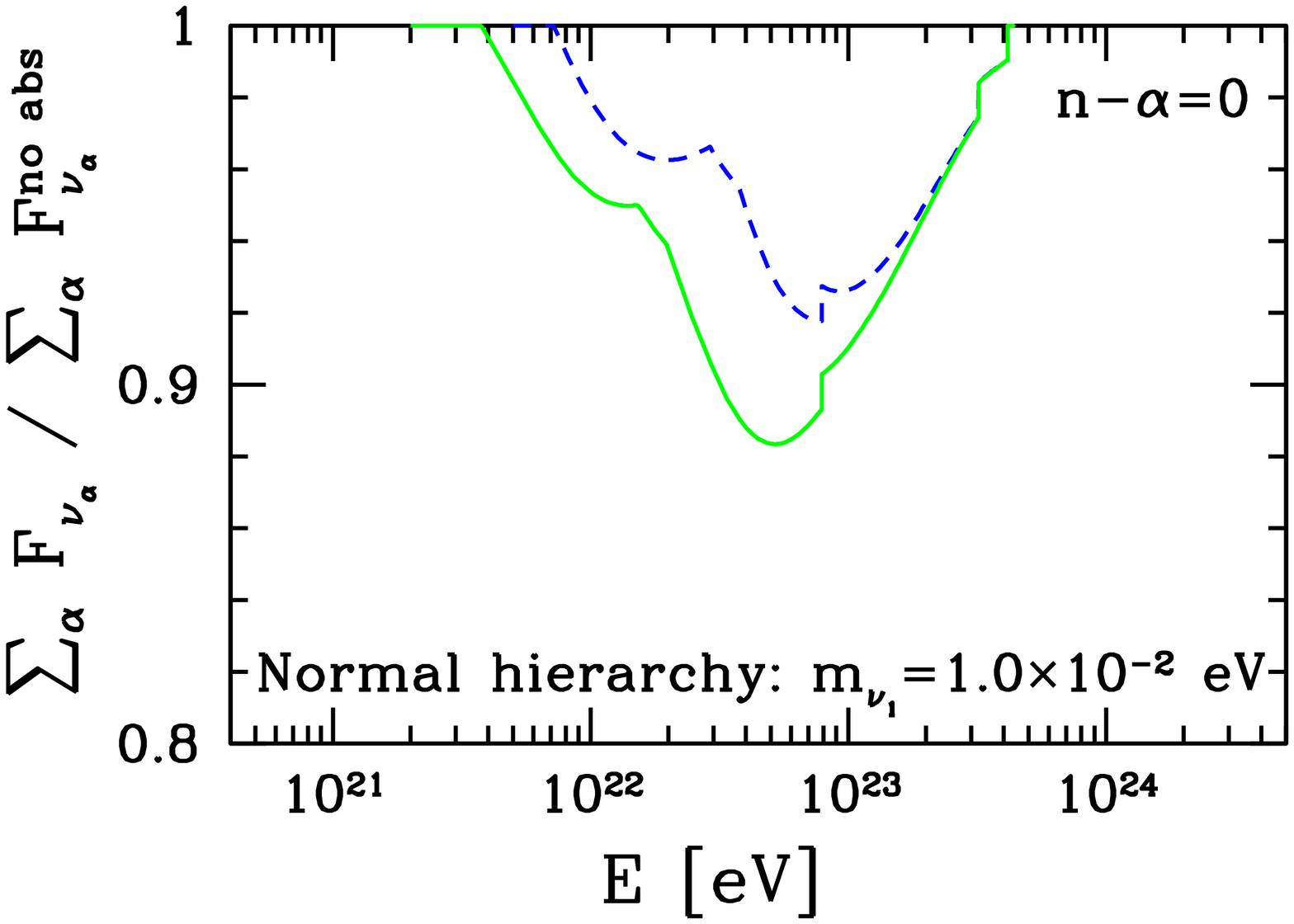}
\includegraphics*[bbllx=20pt,bblly=221pt,bburx=570pt,bbury=608pt,width=7.2cm]{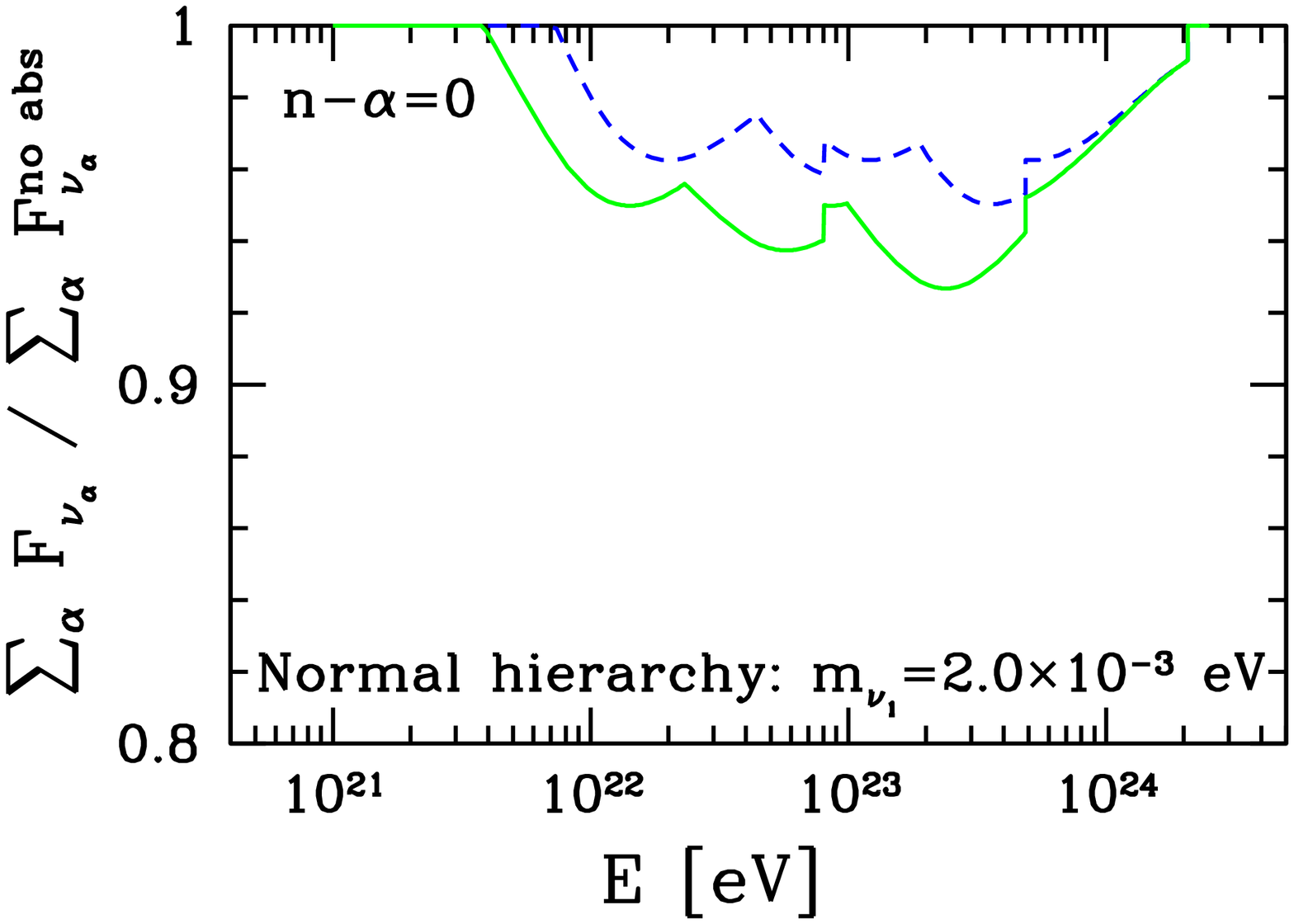}
\includegraphics*[bbllx=20pt,bblly=221pt,bburx=570pt,bbury=608pt,width=7.2cm]{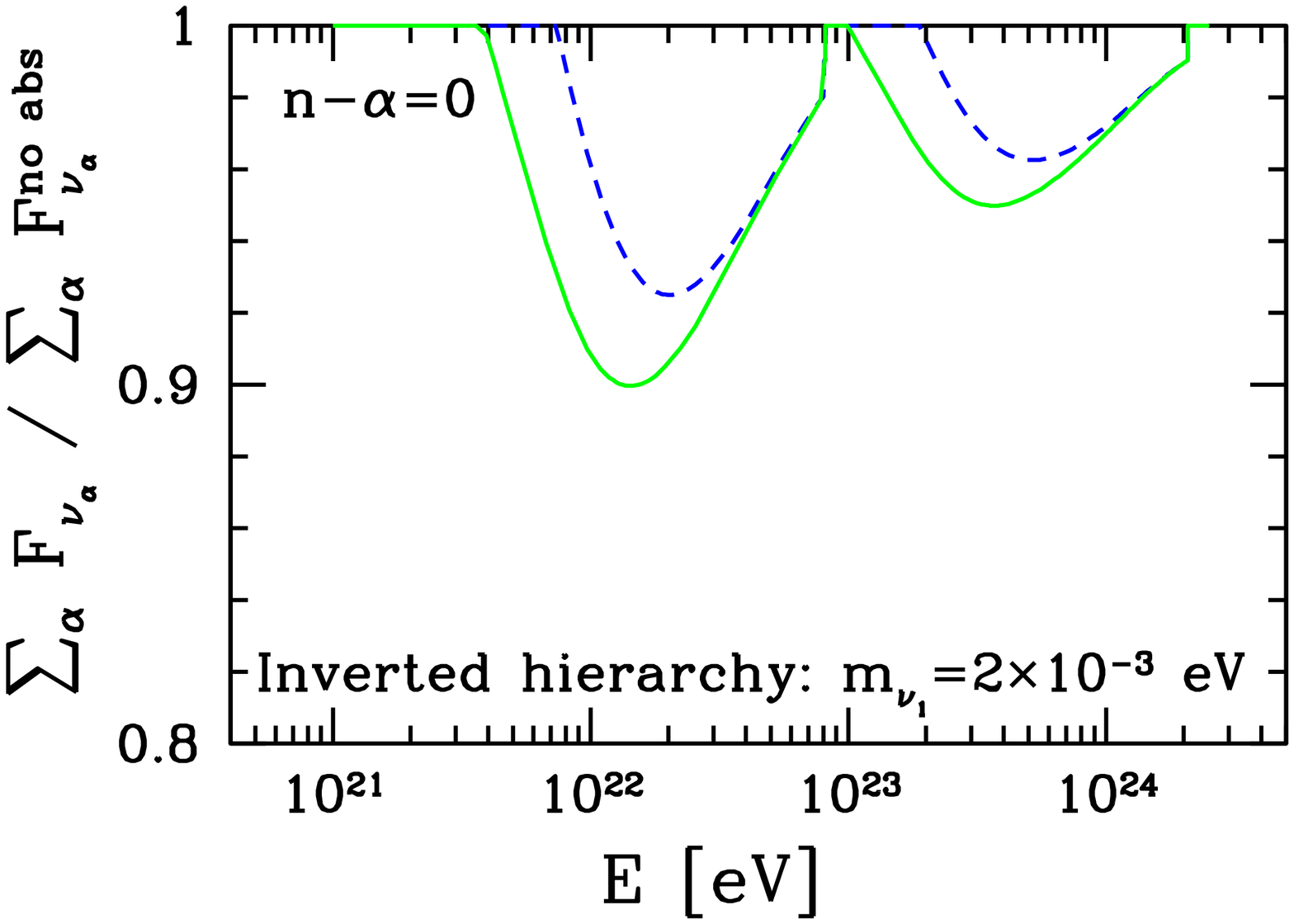}
\caption[...]{As in Fig.~\ref{absorpt-consSFR-norm}, but with a power-law activity~(\ref{activity-pow}) 
and $n-\alpha = 0$, with $z_{\rm min}=0$ and $z_{\rm max} =10$ (short-dashed), 20~(solid), mimicking
a topological defect source scenario.  
For all curves it is assumed that $E_{\rm max}>E_{\nu_1}^{\rm res}\,(1+z_{\rm max})$. 
\label{absorpt-pow-norm-k0}} 
\end{center}
\end{figure}

\begin{figure}
\begin{center}
\includegraphics*[bbllx=20pt,bblly=221pt,bburx=570pt,bbury=608pt,width=7.2cm]{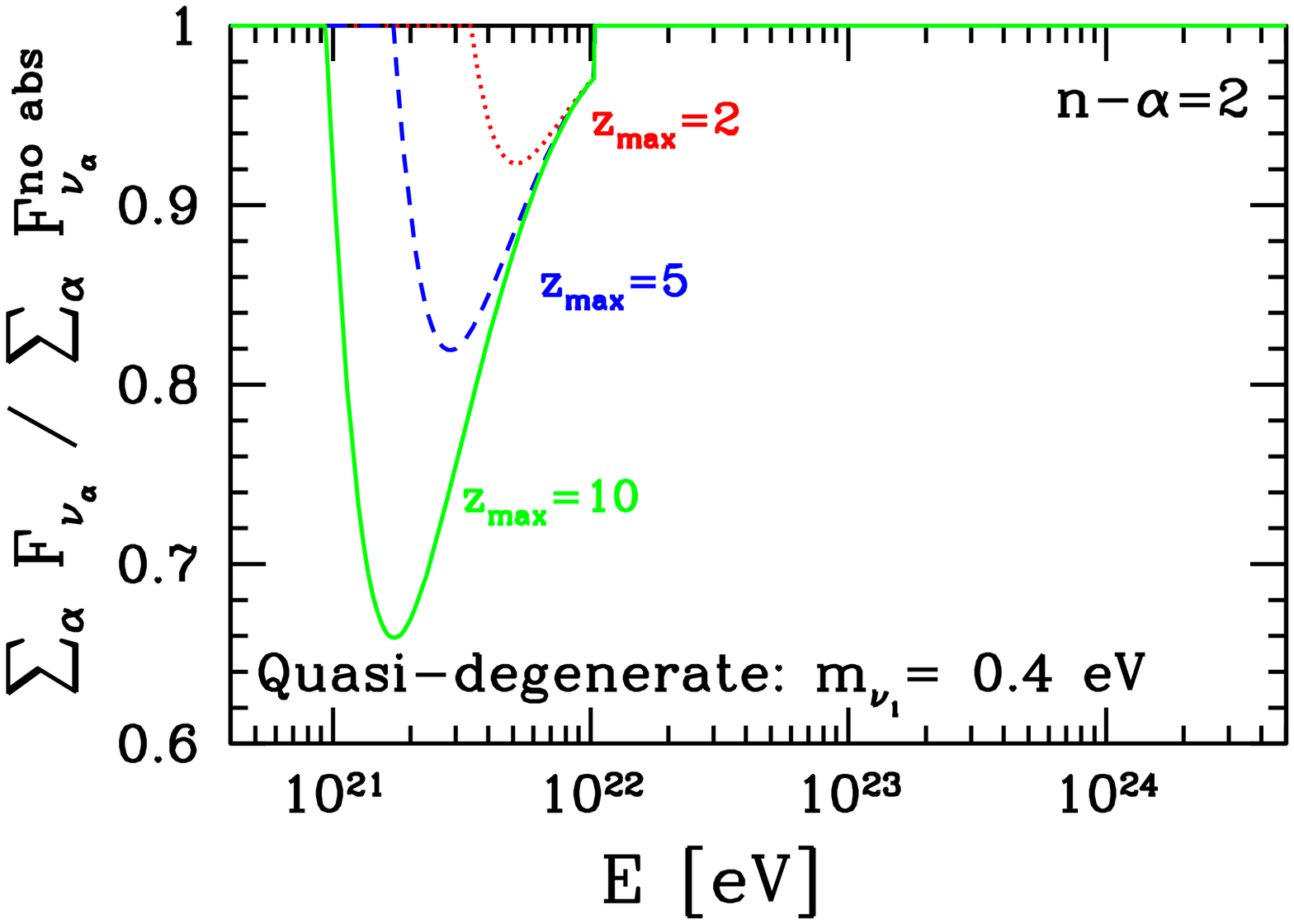}
\includegraphics*[bbllx=20pt,bblly=221pt,bburx=570pt,bbury=608pt,width=7.2cm]{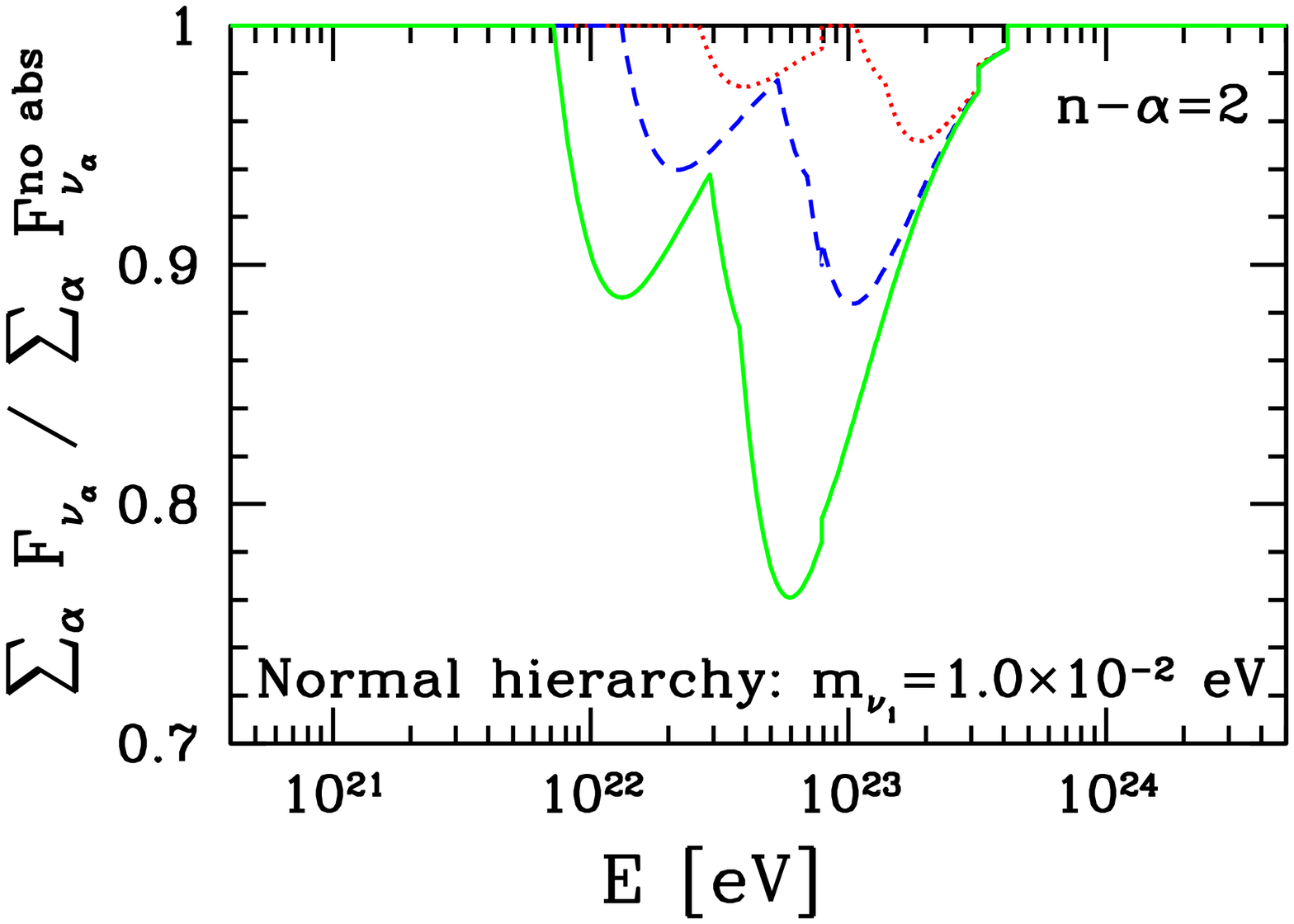}
\includegraphics*[bbllx=20pt,bblly=221pt,bburx=570pt,bbury=608pt,width=7.2cm]{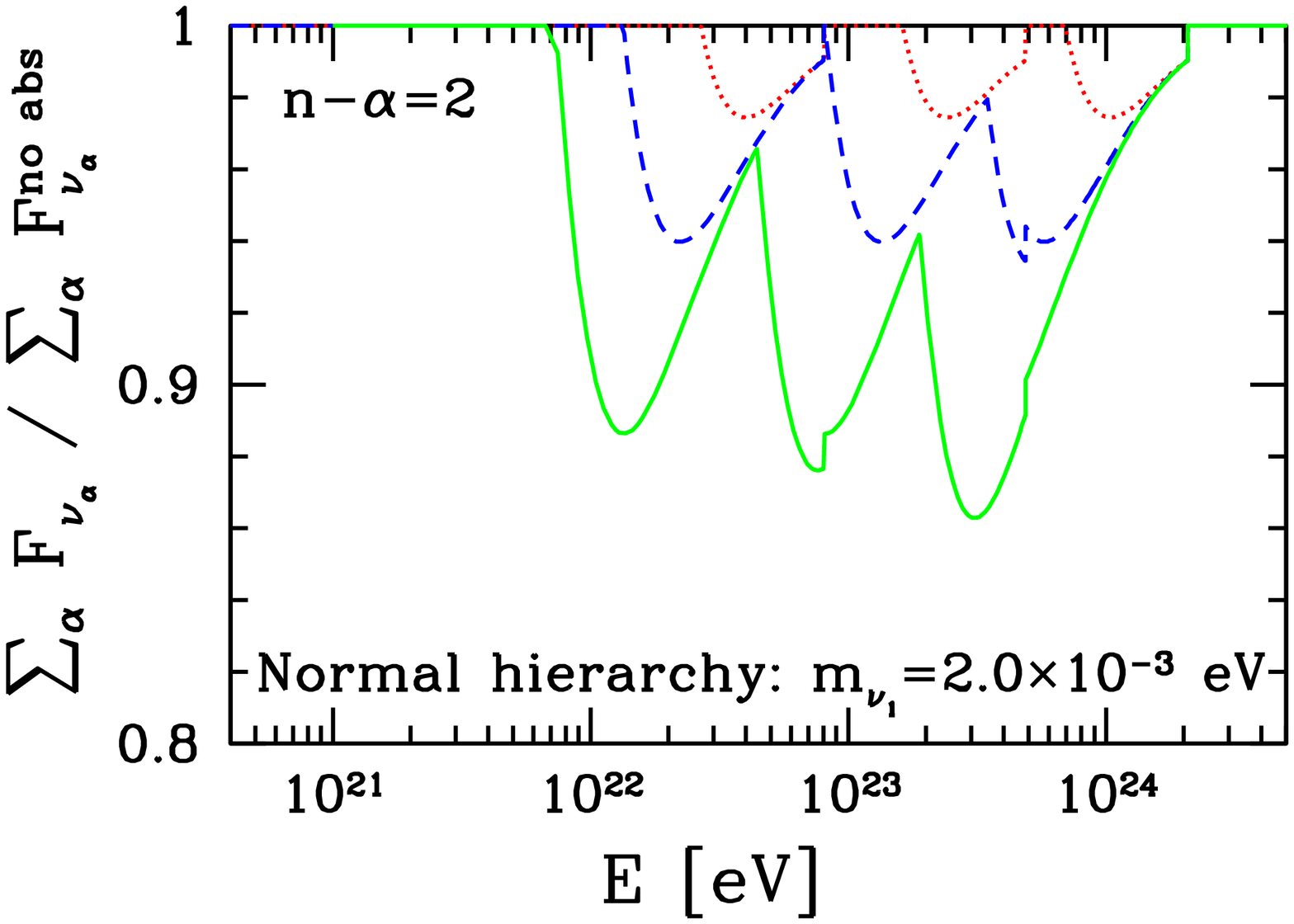}
\includegraphics*[bbllx=20pt,bblly=221pt,bburx=570pt,bbury=608pt,width=7.2cm]{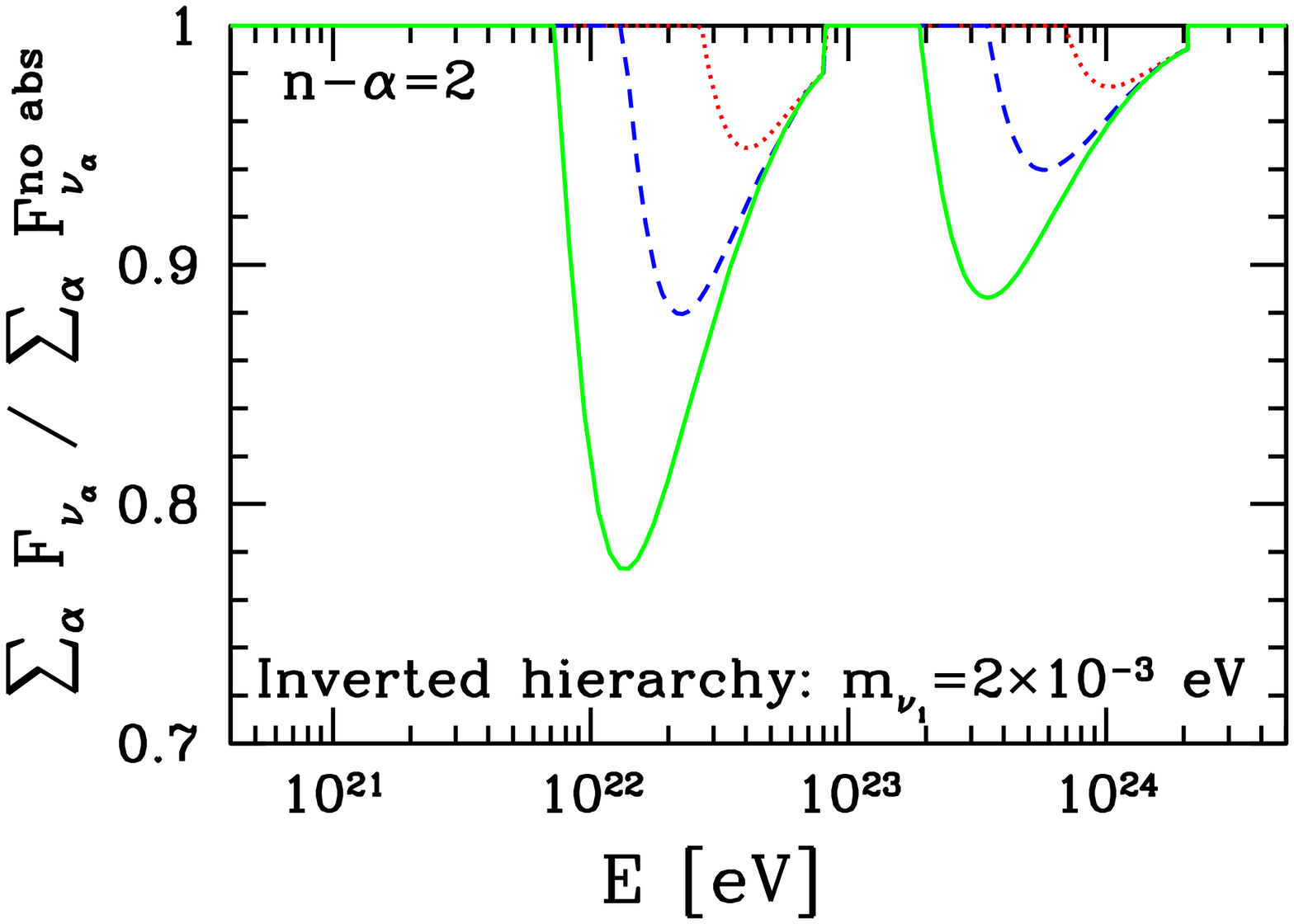}
\caption[...]{Same is Fig.~\ref{absorpt-pow-norm-k0}, but with $n-\alpha = 2$ fixed,
and $z_{\rm max} =2,5,10$ (from upper to lower curves), corresponding to a bottom-up acceleration source scenario.
\label{absorpt-pow-norm}} 
\end{center}
\end{figure}

Next, we consider some non-degenerate neutrino mass scenarios,
and source activities other than power-law. 
In the four panels of Figs.~\ref{absorpt-consSFR-norm}\,--\,\ref{absorpt-pow-norm}, 
we show the prediction~(\ref{flux-earth-appr-120-sum}) for 
$\sum_\alpha F_{\nu_\alpha}/\sum_\alpha F_{\nu_\alpha}^{\rm no\ abs}$ for  
a neutrino spectrum which is quasi-degenerate, i.e. lightest mass $m_{\nu_1}= 0.4$~eV (upper panel),  
normal hierarchical, with $m_{\nu_1} = 0.01$~eV (2nd panel) and $m_{\nu_1} = 0.002$~eV (3rd panel),
and inverted hierarchical with $m_{\nu_1} = 0.002$~eV (bottom panel).
Figures~\ref{absorpt-consSFR-norm} and \ref{absorpt-genSFR-norm} display results for $z_{\rm max}=10$,
$\alpha=2$~(solid) and 0~(dashed), and the conservative and generous star-formation 
activities, respectively.
Figure~\ref{absorpt-pow-norm-k0} shows results for a power-law activity, $n-\alpha=0$, and 
$z_{\rm max}=10$~(short-dashed) and 20~(solid), mimicking a topological defect source scenario. 
Fig.~ \ref{absorpt-pow-norm} displays results for a power-law activity, $n-\alpha=2$, 
and $z_{\rm max}=$~2, 5, 10 from upper to lower curves, corresponding to bottom-up acceleration.  
As expected, 
the depths of the dips increase markedly with $z_{\rm max}$, 
lower spectral index $\alpha$~(\ref{inj-spect-pow}),
and increased source evolution~(\ref{activity-pow}),
the latter represented either by the shift from ``conservative'' to ``generous''
SFR as in Figs.~\ref{absorpt-consSFR-norm} and \ref{absorpt-genSFR-norm},
or by the increase in $n-\alpha$ as in 
Figs.~\ref{absorpt-pow-norm-k0} and \ref{absorpt-pow-norm}. 

The 2nd panels in the figures illustrate that in the case of a normal
hierarchy and an intermediate lightest neutrino mass of $m_{\nu_1}=0.01$~eV, 
corresponding to central values $m_{\nu_2}= 0.013$~eV and 
$m_{\nu_3}=0.053$~eV in Fig.~\ref{mass-scen} (top), 
the absorption dips from $\nu_1$ and $\nu_2$ cannot be resolved.
The two dips appear as one dip, which is understandably 
about twice as deep as the single dip from $\nu_3$.

The two lower panels present results for a very small, lightest neutrino 
mass of $m_{\nu_1}=0.002$~eV.  
In the normal hierarchical case (3rd panel), 
three absorption dips 
-- associated with the 
masses $m_{\nu_1}=0.002$~eV, 
and, from the central values of Fig.~\ref{mass-scen} (top),
$m_{\nu_2}= 0.0085$~eV and 
$m_{\nu_3}=0.052$~eV -- 
are clearly visible. 
These 3rd panels present relic neutrino mass absorption spectroscopy at its best! 
For an inverted hierarchy, on the other hand, one finds, as 
illustrated in the bottom panels,
that the absorption dips from $\nu_2$ and $\nu_3$ 
cannot be resolved individually.
With $m_{\nu_1}=0.002$~eV, the central values for the nearly-degenerate heavy 
masses are $m_{\nu_2}\simeq m_{\nu_3}=0.052$~eV, from 
Fig.~\ref{mass-scen} (bottom).

The 2~meV mass is a critical one in the sense that 
the relic neutrino $\nu_1$ is indeed non-relativistic for $z$ up to $z_{\rm max}=2$,
in which case Eq.~(\ref{relic-dens}) is applicable.  For even smaller masses (or larger
$z_{\rm max}$),  
the relic $\nu_1$ density becomes relativistic and the absorption dip is known to 
wash out~\cite{Weiler:1982qy}.  
Thus, the two lower panels in Figs.~\ref{absorpt-consSFR-norm}\,--\,\ref{absorpt-pow-norm}
remain applicable for any $m_{\nu_1}<0.002$~eV, if one simply removes the contribution of the 
highest-energy dip, at $\sim 10^{24}$~eV.



\section{Experimental prospects\label{prospects}}

In the last paragraph we have seen that, depending on the source activity,
injection spectrum, and neutrino masses, 
we may expect absorption dips of $(10\div 20)$~\% depth, 
located at $(0.1\div 0.5)\,E_{\nu_i}^{\rm res}$, with a width of about 
an order of magnitude in energy.
Is there any hope of discovery of these absorption dips 
in the next decade or beyond? 
The answer to this question depends critically 
on the magnitude of the EHEC$\nu$ flux at the resonance energies realized in Nature.
In the following we will study a few benchmark flux scenarios (\S~\ref{benchmarks}) 
and discuss further experimental (\S~\ref{issues}) issues concerning the 
clean reconstruction of absorption dips.      

\subsection{Benchmark flux scenarios\label{benchmarks}}

\subsubsection{Most optimistic scenario: hidden sources\label{optimistic}}

The most favorable case for relic neutrino absorption spectroscopy is 
the one in which the neutrino fluxes saturate the current observational 
upper bounds. 
It is known quantitatively (cf. the points marked 
``Z-burst''  from Ref.~\cite{Fodor:2001qy} in Fig.~\ref{roadmap} (bottom)) 
that in this most favorable flux scenario, 
the secondary fluxes of protons (and photons) from hadronic Z-decay 
are remarkably of just the right order of magnitude to explain 
the highest energy cosmic rays above the GZK energy
by Z-bursts~\cite{Weiler:1999sh,Fargion:1999ft}. 

As can be seen from Fig.~\ref{roadmap} (bottom) and as summarized in Table~\ref{ev-numb}, 
the upcoming EHEC$\nu$ observatories expect to see in this case 
in 2013 about $230$ neutrinos (plus anti-neutrinos) per flavor $\alpha =e,\mu,\tau$ in the 
energy interval from $10^{21}$~eV to $10^{22}$~eV.
The total number of neutrino events, then, is $N\simeq 700$ neutrinos
of all flavors, which implies a 1-sigma fluctuation of $\sqrt{N}\simeq 26$. 
For a 3-sigma evidence for an absorption dip in this energy interval, 
an absorption depth of $3\,\sqrt{N}/N\simeq 11$~\% is required; 
for a 5-sigma discovery, a depth of  $5\,\sqrt{N}/N\simeq 19$~\% is required. 

As a comparison with our case studies in \S~\ref{case} reveals, 
these depth requirements are achievable as long as the dip is maximized via a quasi-degenerate 
neutrino mass.  
The quasi-degenerate condition is $m_{\nu_1}\gwig\, 0.1$~eV,
which in turn implies that $E_{\nu_i}^{\rm res}\lwig 4\times 10^{22}$~eV. 
For example, power-law source emissivities produce such depths 
for $n-\alpha\,\gwig\, 0$ and/or  $z_{\rm max}\gwig 10$ (cf.\ Figs.~\ref{absorpt-pow-deg}, 
\ref{absorpt-pow-norm-k0} (top), and \ref{absorpt-pow-norm} (top)). 
Therefore, the realization of the Z-burst mechanism 
for the already observed highest energy cosmic ray events implies 
not only a discovery of the corresponding EHEC$\nu$ flux by 2008 
(cf.\ Table~\ref{ev-numb}), but also 
the discovery of solid evidence for the associated absorption dip as soon as 2013. 

\begin{table}
\caption{Expected number of neutrinos (plus anti-neutrinos) to be detected in 
upcoming EHEC$\nu$ observatories with energies 
in the indicated intervals until the indicated year, for two different EHEC$\nu$ flux scenarios 
-- one saturating the current observational upper bound and one saturating the cascade limit 
(cf. Fig.~\ref{roadmap} (bottom)).
\label{ev-numb}}
\begin{ruledtabular}
\begin{tabular}{|l||r|r|r|r|r|r|}
 & \multicolumn{6}{c|}{$\sum_\alpha \triangle\, (N_{\nu_\alpha}+N_{\bar\nu_\alpha})$} 
\\[1.5ex] \hline 
energy decade & \multicolumn{2}{c|}{$10^{21\div 22}$~eV} 
 & \multicolumn{2}{c|}{$10^{22\div 23}$~eV} 
 & \multicolumn{2}{c|}{$10^{23\div 24}$~eV} \\[1ex] \hline
year   & 2008 & 2013 & 2008 & 2013 & 2008 & 2013 \\[1ex] \hline\hline
observ. limit & 240 & 700 & 30 & 90 & 2 & 5 \\ \hline
cascade limit & 13 &40 & 3 &10 & 1 &2 \\
\end{tabular}
\end{ruledtabular}
\end{table}

What kind of source could produce such an optimistic EHEC$\nu$ flux, of the order of
the observational limit at $10^{21}$~eV to $10^{22}$~eV? 
Bottom-up astrophysical Zevatrons may in principle produce such a flux,
with the problem that they must accelerate protons to energies 
$E_{p\, {\rm max}}\gwig 10^{23}$~eV~\cite{Fodor:2001qy,Kalashev:2001sh}. 
Much more favorable in this respect are top-down sources, 
involving physics beyond the Standard Model (SM) which naturally produce 
extremely-high energetic particles.  
With either bottom-up or top-down, the sources must  be ``hidden''~\footnote{ 
Possible astrophysical hidden sources have been discussed in 
the textbook~\cite{Berezinsky:book} and in Ref.~\cite{Berezinsky:2000bq}.
}, in order to avoid the cascade limit (cf. Fig.~\ref{roadmap} (bottom)).  
Hidden sources are, by definition, sources from which neither nucleons nor photons 
(with energies $\gwig\, 100$~MeV) escape.

\begin{figure}
\begin{center}
\includegraphics*[bbllx=20pt,bblly=221pt,bburx=585pt,bbury=626pt,width=8.65cm]{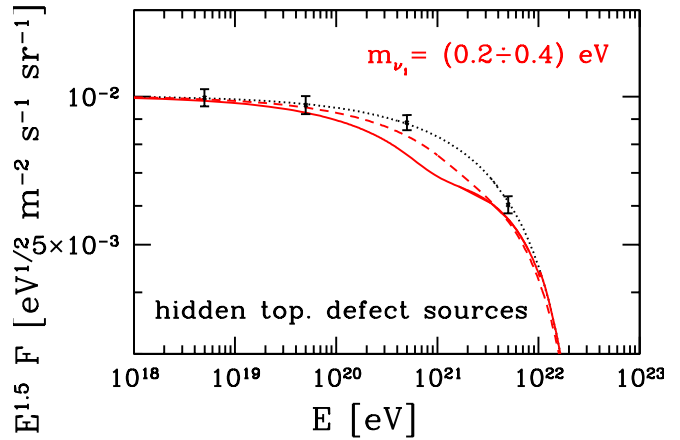}
\caption[...]{Predicted neutrino flux at Earth, summed over all flavors, 
from a power-like source emissivity, 
with $n=1.5$, $z_{\rm max}=\infty$, $\alpha =1.5$, 
and $E_{\rm max}=4\times 10^{22}$~eV. 
This flux mimics one from hidden-sector topological defects with 
$M_X=4\times 10^{14}$~GeV  (cf.\ Fig.~\ref{roadmap} (bottom)) and is also sufficient 
to explain the EHECR's above $E_{\rm GZK}$ via the Z-burst mechanism. 
Curves are without (dotted) and with relic neutrino absorption.  
Assumed neutrino masses are degenerate at $m_{\nu_1}=0.2$~eV (dashed) and $m_{\nu_1}=0.4$~eV (solid).
The error bars indicate the statistical accuracy achievable 
per energy decade by the year 2013, for a flux which saturates today's observational bound from 
Fig.~\ref{roadmap} (bottom). 
\label{sign_hidden}} 
\end{center}
\end{figure}

One popular hidden top-down example is a topological defect,
here taken with mass $M_X\sim 4\times 10^{14}$~GeV, 
which couples to SM particles only indirectly through 
a non-SM sector~\cite{Berezinsky:1999az}.  
In this case of topological-defect origin, 
we have $n-\alpha\simeq 0$ and $z_{\rm max}\gg 1$, 
and so Fig.~\ref{absorpt-pow-norm-k0} applies;
the dip presented there is $\sim 15$~\% for the most interesting
case of quasi-degenerate neutrinos, $m_{\nu_1}\gwig\, 0.1$~eV.
In Fig.~\ref{sign_hidden}, we show the significant wiggle from quasi-degenerate neutrinos 
in the otherwise power-law spectrum
expected for the hidden topological-defect neutrino flux as in Fig.~\ref{roadmap} (bottom).  
By design, this flux 
scratches the observational upper bound. 
Such a flux is also sufficient to explain the EHECR's via the Z-burst mechanism.
The indicated error bars show the statistical significance that is expected 
with planned and proposed experiments by the year 2013 (cf. Fig.~\ref{roadmap} (bottom)).   
This example illustrates concretely our 
previous claim that substantial evidence for a relic neutrino absorption dip 
and the existence of the C$\nu$B is achievable within the 
foreseeable future, if the Z-burst mechanism for the EHECR's is realized 
in Nature. 

For a hierarchical neutrino spectrum, on the other hand, with $m_{\nu_1}\lwig\, 0.04$~eV,
the lowest resonant-energy is $\sim 10^{23}$~eV.
Even if the EHEC$\nu$ flux saturates the current observational limit,
the expected event numbers in the $10^{22\div 23}$~eV energy interval 
are apparently too small (cf.\ Table~\ref{ev-numb})
to allow a discovery of absorption dips in the next decade: 
with 90 expected events in this energy range by 2013,
one needs absorption depths of $\sim 53$~\% ($\sim 32$~\%)
for a 5-sigma discovery (3-sigma evidence).
As can be gleaned from Fig.~\ref{absorpt-pow-norm-k0}, 
such large dips are not expected.

\begin{figure}
\begin{center}
\includegraphics*[bbllx=20pt,bblly=221pt,bburx=585pt,bbury=626pt,width=8.65cm]{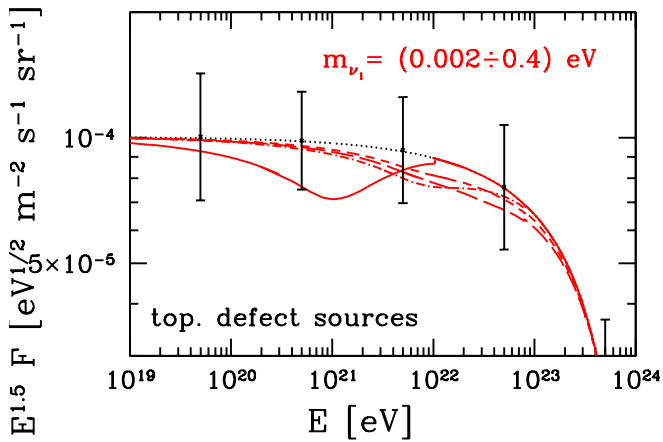}
\caption[...]{As in Fig.~\ref{sign_hidden}, but with 
$E_{\rm max}=10^{24}$~eV, to 
mimic  topological defects with $M_X=10^{16}$~GeV (cf.\ Fig.~\ref{roadmap} (bottom)).
The assumed neutrino spectra are: (i) quasi-degenerate, $m_{\nu_1}=0.4$~eV (solid), 
(ii) normal hierarchical, $m_{\nu_1}=0.01$~eV (long-dashed) and 
$m_{\nu_1}=0.002$~eV (short-dashed), and 
(iii) inverted hierarchical, $m_{\nu_1}=0.002$~eV (dashed-dotted).
Here the error bars indicate the statistical accuracy achievable 
per energy decade by the year 2013, for a flux which saturates today's cascade limit from 
Fig.~\ref{roadmap} (bottom).
\label{sign_casc}} 
\end{center}
\end{figure}

\subsubsection{Less optimistic scenario: transparent  sources\label{less-optimistic}}

Let us turn our attention now to the non-optimized EHEC$\nu$ fluxes. 
From a neutrino flux that saturates the cascade limit (cf.\ Fig.~\ref{roadmap} (bottom)), 
we may expect, in the $10^{21\div 22}$~eV energy bin, just 40 events by the year 2013 
(cf.\ Table~\ref{ev-numb}).
This yields a 3-sigma evidence for an absorption
dip only if it has a depth of $\sim 48$~\%~\footnote{
If the number of events are small, one should apply Poisson statistics, as given in 
Ref.~\cite{Gehrels:mj}.}.  
Dips of this depth we have seen in \S~\ref{case} only for 
quasi-degenerate neutrinos and 
extreme parameter choices for the source activity.
For example, for power indices $n-\alpha\gwig\, 2$, 
as might happen with highly-evolving bottom-up sources,
one needs $z_{\rm max}\gwig\, 20$ in addition 
(cf.\ Figs.~\ref{absorpt-pow-deg} and \ref{absorpt-pow-norm} (top)).
The latter is easily achieved in top-down sources ($z_{\rm max}=\infty$), however their restriction to 
$n-\alpha\simeq 0$ tends to decrease the depth of the dip 
(cf.\ Figs.~\ref{absorpt-pow-deg} and \ref{absorpt-pow-norm-k0} (top)).
This pessimistic outlook is however ameliorated when one realizes 
that an increase in statistics by a factor of $10$ reduces the required absorption depth
by a factor of $\sim 3$, to $\sim 15$~\% for a 3-sigma evidence, 
and to $\sim 25$~\% for a 5-sigma discovery. 
Such an increase in statistics could be achieved, for example, 
by undertaking more ANITA flights, by extending the EUSO flight time,
or by developing the OWL or SalSA experiments. 
If the $\sim 13$ neutrinos in the $10^{21\div 22}$~eV interval 
(cf.\ Table~\ref{ev-numb}), 
expected from a flux which saturates the cascade limit,
are measured by the pre-2008 experiments,
then such extensions of the upcoming EHEC$\nu$ observatories would be warranted.  

\begin{figure}
\begin{center}
\includegraphics*[bbllx=20pt,bblly=221pt,bburx=585pt,bbury=626pt,width=8.65cm]{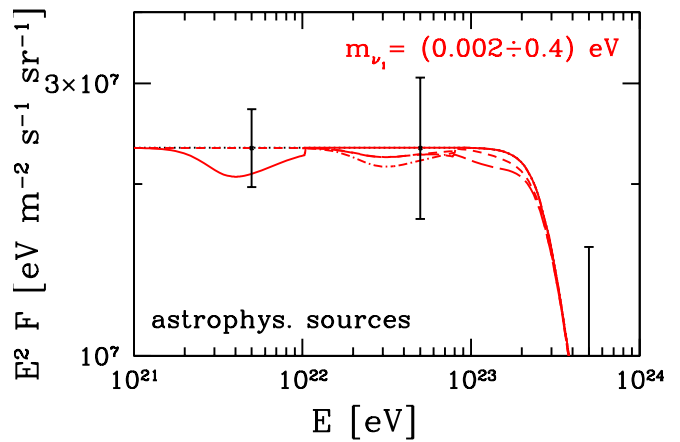}
\caption[...]{As in Fig.~\ref{sign_casc}, but with 
a generous SFR activity~(\ref{activity-SFR}), $z_{\rm max}=20$, injection spectrum index 
$\alpha =2$, and $E_{\rm max}=10^{24}$~eV, to 
mimic  astrophysical sources with $E_{p\,{\rm max}}=2\times 10^{16}$~GeV.
\label{sign_casc_astro}} 
\end{center}
\end{figure}

What classes of sources could deliver EHEC$\nu$ fluxes which saturate the cascade 
limit? It has recently been pointed out that cosmogenic neutrino fluxes
from photo-pion production by cosmic protons on the CMB can reach this 
limit~\cite{Fodor:2003ph,Kalashev:2002kx}.
However, the limit is scratched typically just above $E_{\rm GZK}$, at around $10^{20}$~eV 
and not higher (cf.\ Fig.~\ref{roadmap} (bottom)). 
Other possible sources include bottom-up Zevatrons
with a large $E_{p\,{\rm max}}\gwig 10^{23}$~eV, 
and topological defects with $M_X\gwig 10^{14}$~GeV.

As examples, we show in Figs.~\ref{sign_casc} and \ref{sign_casc_astro}, 
for various neutrino mass spectra, the predicted modulations in the otherwise power-law spectra 
expected for a topological defect and an astrophysical neutrino flux, respectively, both saturating 
the cascade limit.  
As expected, we see that a further increase in statistics by a factor of $\sim 10\div 100$, to 
reduce the error bars by a factor of $\sim 3\div 10$, will be mandatory to establish absorption dips.
Moreover, a quasi-degenerate neutrino mass spectrum is required.

\subsection{Further experimental issues\label{issues}}

To temper optimism, it is fair to mention three further issues that 
mitigate an experiment's ability to cleanly observe an absorption dip.
These are (i) the resolution with which an experiment can reconstruct
the initial neutrino energy from the visible event energy,
(ii) the flavor-dependent nature of energy reconstruction,
and (iii) the possible confusion of a deviation of the (assumed) power-law
spectrum due to absorption with other possible origins of deviation.
We discuss each of these in turn.

Proposed EHEC$\nu$ detectors will measure shower energy 
from ground-based scintillator or water, or atmospheric 
nitrogen fluorescence, or radio signals in ice.
From EHECR physics, energy reconstruction for hadron-initiated showers 
is known to be about 25~\%.
Neutrino observatories expect to reconstruct shower energies with about the 
same error. This will smear the dip somewhat,
but our analysis is based upon events per decade in energy, so the 25~\% 
smearing should be tolerable.

However, there is a subtlety connected with the various showers that 
result from different neutrino flavors~\footnote{
We take our numbers from the extrapolated cross-sections in~\cite{GQRS}.
}.
Consider first neutral current (NC) events,
identical for all neutrino flavors.  
On average, the final state neutrino carries away 80~\% of the incident energy,
leaving 20~\% in the detected jet.
The NC cross-section is $\sim 45\%$ that of the charged-current (CC) cross-section.
Next consider CC interactions.  The
$\numu$ and $\nutau$ CC events produce charged leptons,
generally unobservable,
which carry away 80~\% of the energy.
Thus, the $\numu$ and $\nutau$ CC events 
leave the same 20~\% energy deposition in the shower 
as do the NC interactions.
For all these events, the absorption dip in the shower events will appear 
lower in energy than the true dip by a factor $\sim 0.2$.
This is easily corrected.  However, there remains some smearing due to the  
event-by-event variance of the energy transfer about the mean.
We do not include the effect of this variance in the present work.

In contrast to the above event classes, a $\nue$ CC event will produce 
a hadronic jet plus an electron which creates an electromagnetic jet,
and so the interaction deposits 100~\% of the incident energy 
into the combined shower.  

In summary of this flavor discussion, NC, and $\numu$ and $\nutau$ CC 
scattering constitute about 77~\% of the events.
The showers from these events contain about 20~\% of the incident 
neutrino energy. 
The other 23~\% of events are due to $\nue$ CC,
with all of the incident energy observed.
Thus, the observed events will look like a superposition of two fluxes
with relative weights 23~\% and 77~\%,
the latter displaced downward in energy by a factor of five in the 
mean, but with fluctuations.

Finally we come to the confidence issue, 
whether an observed feature near the end of the EHEC$\nu$ spectrum can be claimed to be an absorption dip.
For example, the spectral end-point may have structure simply due to differences
among the individual contributing sources.
To better ensure that a dip feature is observed,
the continuation of the spectrum {\sl above} the dip region 
should be observed.  This requires more events, at higher energies still.
Looking again at Table~\ref{ev-numb}, one sees even more reason to doubt the 
observation of an absorption feature, except for relatively low resonance-energies.
In turn, this means that quasi-degenerate neutrinos, 
with mass not much below present cosmological bounds,
are required if the relic neutrinos are to be detected via an 
absorption dip.

\section{Summary and Conclusions\label{conclusions}}
 
Relic neutrino absorption spectroscopy via 
the observation of absorption dips in the EHEC$\nu$ flux 
may be feasible. 
The Z-dips, if observed,
are rich in particle and astrophysical information.
The position of the high-energy edge of each dip is fixed by the neutrino mass.
The depth and shape are determined by the density of the C$\nu$B, by the usual cosmological parameters, 
by the activity and injection spectrum of the EHEC$\nu$ sources, and
by the neutrino mass pattern. 
Moreover, as we have seen, the width of the dip region 
reflects the spectrum and evolutionary history of the neutrino source(s).
Thus, were we so fortunate as to resolve a dip in some detail,
the information to be mined truly belongs to interdisciplinary particle-astrophysics.

However, the statistics at the dip is entirely determined by the magnitude of the 
EHEC$\nu$ flux. Moreover, the energy positions of the dips 
-- and, to some extent, their depths -- critically depend
on the neutrino mass spectrum. 

Large event samples, $N\gg 100$, beyond 
$10^{21}$~eV are needed to reveal Z-dips with statistical significance. 
To get these event numbers within the next decade or so, an EHEC$\nu$ flux   
at least as large as the present cascade limit is required. 
Almost certainly, even higher fluxes must be invoked. They could be generated 
by hidden sources, which are opaque to nucleons and high-energy photons, thereby evading the 
cascade limit.  
Such sources are likely indicative of new physics.
Moreover, the high flux requirement implies a Z-burst contribution 
to the EHECR events beyond $E_{\rm GZK}$ within an order of magnitude of the 
present AGASA rate. The Auger project should detect these large EHEC$\nu$ and 
EHECR fluxes within a few years, and EUSO should easily measure them within the decade.
 
As inferred from our numerous figures, a
quasi-degenerate neutrino mass spectrum with $m_{\nu_1}\gwig\, 0.1$~eV
seems to be required to produce a detectably-deep absorption dip at a 
possibly-accessible resonant energy.  
Such a spectrum is testable in several ways~\cite{Paes:2001nd}.
Neutrinoless double beta decay experiments (assuming neutrinos are Majorana)
such as CUORE~\cite{CUORE} and NEMO-3~\cite{NEMO},    
and the KATRIN tritium decay experiment~\cite{KATRIN} 
can be expected to show positive results already in the upcoming decade.
The transfer functions relating CMB fluctuations to today's large-scale matter distributions 
can be expected to show positive contributions from massive neutrinos~\cite{Hu:1997mj}. 
It seems sensible to say that, if pre-2008 experiments do not see any EHEC$\nu$ flux
in the $10^{21\div 23}$~eV region, 
then, in the context of the concordance cosmological model, 
absorption dips won't be observed within the next decade or two. 
If such is the case, then Nature will have overlooked a wonderful opportunity 
to produce direct evidence for the C$\nu$B~\footnote{
Even if Z-dips cannot be measured, 
it may still be possible to infer the C$\nu$B from Z-burst data.
The statistics of ``emission'' spectroscopy are not as formidable as those of 
absorption spectroscopy.  In emission spectroscopy well above $E_{\rm GZK}$,
each event is background free, and therefore statistically significant.\\
Finally, let us comment on a possible 
loophole to greater event rates than those expressed so far. 
We have worked in the context of the concordance model of 
cosmology.  There are some chinks in the armor.  It has been known for some time that 
simulations with concordance parameters over-produce small-scale objects 
(dwarf galaxies and satellites) compared to observation~\cite{satellites}.  
More-recent observational evidence suggests that the  largest scales may 
also conflict with the concordance model.  There is some evidence that elliptical galaxies, 
galactic clusters, and even vast filaments and walls delimiting huge voids 
may have formed very early in the universe ($z\geq 2$)~\cite{earlymassives}.
Such precocious structure suggests top-down hierarchical development rather than bottom-up as 
predicted by the concordance model.   
The parameters of the concordance cosmology may give way to something new.
All of this encourages an open mind.  
If large structures did form early,
then more neutrino sources and strong source evolution 
may be the reality.  With precocious clustering of matter, precocious clustering of 
neutrinos also becomes possible.
Neutrino clustering in our Galactic Supercluster would enhance ``local'' absorption,
and thereby create narrower and deeper Z-dips (it would also greatly enhance 
the local Z-burst rate).}.

In summary, the presently planned neutrino detectors
open up a window of opportunity for relic neutrino
absorption spectroscopy. The next decade will be
really exciting and decisive in this respect.

\vspace{2ex} 

\appendix*
\section{Implementing Neutrino Flavor Physics}

At production in the cosmic sources,
let the ratios of the neutrino flavors be written as 
$\nue:\numu:\nutau=w_e:w_\mu:w_\tau$, with $\sum w_\beta=1$.
A convenient description of the flavor mixture is given by the
density matrix%
\beq{rho1}
\rho(t=0)=
\sum_\beta w_\beta\; |\nu_\beta\rangle\langle\nu_\beta|\,.
\eeq
The density matrix is properly normalized to 
${\rm Tr}(\rho(0)=1$, and so describes the ensemble-averaged, single neutrino.

The relic neutrinos, with thermal energies $\sim 3\,kT_\nu\sim 0.5$~meV in the today's epoch,
long ago decohered into their mass-eigenstates~\cite{Weiler:Ringberg}.
Therefore, rewriting the density matrix in the mass-basis allows us to simply include the 
losses due to resonant absorption.
Making use of the mixing matrix notation
$|\nu_\beta\rangle=U_{\beta j}\,|\nu_j\rangle$,
or equivalently,
$U_{\beta j}= \langle\nu_j|\nu_\beta\rangle
        =\langle\nu_\beta|\nu_j\rangle^*$,
the forward-propagated density matrix in the mass basis, 
with resonant absorption, is:
\bwide
\bea{rho2}
\rho(t)&=&\sum_\beta w_\beta\,\sum_j
        \left[{\rm e}^{-{\rm i}\frac{m_j^2}{2E}\,t}\,{\rm e}^{-\frac{\Gamma_j}{2}\,t}\,
        U_{\beta j}\,|\nu_j\rangle\right]\,
        \sum_k
        \left[\langle\nu_k|\,U_{\beta k}^*\,
        {\rm e}^{+{\rm i}\frac{m_k^2}{2E}\,t}\,{\rm e}^{-\frac{\Gamma_k}{2}\,t}\right]\nn\\
        &=&\sum_\beta w_\beta\,\sum_{j,\,k} 
        U_{\beta j}\,U_{\beta k}^*\;
        {\rm e}^{+{\rm i}\triangle m^2_{kj}\,t}\;{\rm e}^{-\frac{\Gamma_j+\Gamma_k}{2}\,t}\;
        |\nu_j\rangle\,\langle\nu_k|\,.
\eea
\ewide
Here, 
$\Gamma_j=c\,n_j\,\sigma_j$ is the annihilation rate, and 
the factor $e^{-\frac{\Gamma}{2}\,t}$ correctly accounts for 
absorption at the amplitude level~\footnote{The possible extermination of 
Schr\"odinger's cat is described in the same way.}. 
As written, this formula does not include energy losses due to redshifting;
however, it is straightforward to incorporate redshifting in the final expressions.

At Earth, the probability to detect flavor $\alpha$ is then,
\bwide
\bea{Pbeta1}
P_{\nu_\alpha\,{\rm detected}}
        =\langle\nu_\alpha|\,\rho(t)\,|\nu_\alpha\rangle =
        \sum_\beta w_\beta \sum_{j,\,k}\; 
        U_{\beta j}U_{\beta k}^*U_{\alpha k}U_{\alpha j}^*\,
        {\rm e}^{-{\rm i}\triangle m^2_{kj}\,t}\;{\rm e}^{-\frac{\Gamma_j+\Gamma_k}{2}\,t}\;.
\eea
\ewide
This probability expression naturally divides into a sum diagonal in the 
mass, and a sum of interfering mass terms from which oscillations arise.
After a bit of algebra, the result is:
\bwide
\bea{Pbeta2}
P_{\nu_\alpha\,{\rm detected}}
        &=&\sum_\beta w_\beta 
        \sum_j |U_{\alpha j}|^2\,|U_{\beta j}|^2\,{\rm e}^{-\Gamma_j\,t}\\
        \nonumber 
   &+&2\,\sum_\beta w_\beta \sum_{j<k}\,{\rm e}^{-\frac{\Gamma_j+\Gamma_k}{2}\,t}\,
        \left[
        \Re(U_{\beta j}U_{\beta k}^*U_{\alpha k}U_{\alpha k}^*)\,
        \cos\left(\frac{\triangle m^2_{kj}\,t}{2E}\right)
        -\Im(U_{\beta j}U_{\beta k}^*U_{\alpha k}U_{\alpha j}^*)\,
        \sin\left(\frac{\triangle m^2_{kj}\,t}{2E}\right)
        \right]\;. 
\eea
\ewide
Setting the annihilation rates to zero returns the usual formula for 
neutrino oscillations (cf., e.g., Ref.~\cite{Giunti:2003qt}). 

Although the neutrino wave-function at extreme high-energies may remain coherent over a cosmic distance,
the oscillating terms are effectively averaged away in any 
realistic observation.  Let us look first at the coherence of the neutrino wave-function.
The spread in the neutrino mass-eigenstates after travel through a distance $D$ 
results from the difference in the group velocities $\beta = \delta E/\delta p$:
%
\bea{spread}
D_{\rm spread}&=&
ct\,\delta\beta = 
\frac{D\,\triangle m^2}{2E^2}\\ \nonumber
        &=&6\times 10^{-20}\,
\left( \frac{0.7}{h}\right) \left(\frac{D}{D_H}\right)
        \,\triangle m^2_{-3}\,E_{22}^{-2}\;{\rm cm}\,,
\eea
%
where $\triangle m^2_{-3}\equiv\triangle m^2/10^{-3}{\rm eV}^2$,
$E_{22}\equiv E/10^{22}$~eV, and
the Hubble distance $D_H\equiv c H^{-1}_0=4.2\,(0.7/h)$~Gpc
characterizes the typical cosmic distance. 
Decoherence of the wave packet results when $D_{\rm spread}$
exceeds the natural length of the wavepacket, call it $c\,\tau_{\Psi}$,
the latter being determined by conditions at the source~\footnote{
There is also a quantum mechanical spreading of the wavepacket,
governed by the phase velocity $E/p$.
This spreading is smaller than the eigenstate separation 
by the tiny factor $\delta p_\Psi/E\sim (c\,\tau_\Psi\,E)^{-1}$~\cite{Kayser},
and therefore negligible.
}.
The natural length may be $c$ times the production time~\cite{Nussinov}
(e.g.\ $c\,\tau_{\pi^\pm}\sim 3$~m),
the mfp between interactions in a dense source~\cite{Nussinov},
or the spatial uncertainty in the location of the production point 
in the source~\cite{Kayser}.
The decoherence length is obtained by setting $D_{\rm spread}$ in Eq.~\rf{spread} 
equal to $c\,\tau_{\Psi}$, and solving for $D$.  
The result, as a fraction of the Hubble size, is
%
\beq{decohere}
\frac{D_{\rm decohere}}{D_H}= 0.5\,\left(\frac{h}{0.7}\right)\,
        \left(\frac{\tau_{\Psi}}{3\,{\rm m}}\right)\,
        \frac{E_{22}^2}{\triangle m^2_{-3}}\times 10^{20}\,.
\eeq
%
Clearly, decoherence does not occur in our Universe 
for a $10^{22}$~eV neutrino.

Now we turn to the averaging effects of measurement.
To observe the oscillating terms requires measurement of the phase 
$\phi=\triangle m^2\,t/2E$ to better than $2\pi$,
or equivalently, knowledge of $\delta(D/E)$ to better than
$4\pi/\triangle m^2$.
This in turn requires knowledge of $\delta D$ to better 
than $\lambda_{\rm osc}\equiv 4\pi\,E/\triangle m^2$,
{\sl and} knowledge of $\delta E$ to better than $E\,(\lambda_{\rm osc}/D)$.
With the oscillation length being so short $\sim \,E_{22}/\triangle m^2_{-3}$~kpc
compared to cosmic scales,
there is no hope to observe the oscillating term.

From here on, we may use for the 
detection probability just the averaged value of Eq.~\rf{Pbeta2} , i.e.\
just the first term
\beq{Pbeta3}
P_{\nu_\alpha\,{\rm detected}}
        =\sum_j |U_{\alpha j}|^2\;e^{-\Gamma_j\,t}\,
        \sum_\beta w_\beta \,|U_{\beta j}|^2\,.
\eeq
Finally, taking into account energy losses due to redshift, Eq.~\rf{Pbeta3} 
is easily generalized to Eq.~(\ref{neutrino-prop}) in the main text 
for the differential propagation function of neutrinos, 
and to Eq.~(\ref{flux-earth-appr}) for the neutrino flux to be observed at Earth. 

As simple as Eq.~\rf{Pbeta3} is,
further simplifications are possible.
For example, if $w_e:w_\mu:w_\tau=1:1:1$,
as might arise from ``democratic'' neutrino 
emission from a topological defect,
then by unitarity of the mixing matrix,
$\sum_\beta w_\beta\,|U_{\beta j}|^2 = 1/3$
independently of $j$.
It is also known~\cite{Learned:1994wg} that the same result obtains for the case 
$w_e:w_\mu:w_\tau=1:2:0$ as results from charged pion decay to 
neutrinos, in the limit of $\numu\leftrightarrow\nutau$~interchange-symmetry.
The latter is exact if 
the atmospheric mixing angle $\theta_{32}$ is maximal ($45^\circ$)
and $\Re(U_{e3})=0$.
Oscillation data~\cite{Hagiwara:fs} show that these two conditions are satisfied (or nearly so).
Since any linear combination of the two sets $\{w_\beta\}$ just 
discussed must also give the same conclusion,
we arrive at a mini-Theorem:\\
whenever $w_e =1/3$ and $\numu\leftrightarrow\nutau$~interchange-symmetry
is (nearly) valid,\\ 
then $\sum_\beta w_\beta\,|U_{\beta j}|^2 = 1/3$ independently  of $j$. \\
Because $w_e =1/3$ includes two popular neutrino-production cases, 
charged pion decay and democratic emission,
we will assume that $w_e =1/3$ holds and work with the the very compact
result
\beq{Pbeta4}
P_{\nu_\alpha\,{\rm detected}}=\frac{1}{3}\,
        \sum_j |U_{\alpha j}|^2\,e^{-\Gamma_j\,t}\,.
\eeq
This expression leads immediately to the main text's 
Eq.~(\ref{flux-earth-appr-120}) for the
neutrino flux at Earth, after again taking into account the energy loss due to redshifting.

Cosmic-neutrino detectors are unlikely to be capable of 
neutrino flavor identification (see Ref.~\cite{Beacom:2003nh} for a discussion).
Accordingly, it is tempting to also sum over $\alpha$, the
flavor arriving at Earth, to arrive at the especially simple result
\beq{Pany}
P_{\nu_{\rm any}}=
\sum_\alpha P_{\nu_\alpha\,{\rm detected}} =
\frac{1}{3}\,\sum_j \;e^{-\Gamma_j\,t}\,,
\eeq
leading to Eq.~(\ref{flux-earth-appr-120-sum}). 
Caution is warranted, however, as the interactions of the different flavors 
deposit different mean energies into the detector.
For example, the short mean-free-path (mfp) of a high-energy electron,
and the long mfp's of high-energy muons and taus, 
imply that in charged-current interactions, $\nu_\mu$'s and $\nu_\tau$'s 
will leave much less visible energy in the detector than will $\nu_e$'s.
This issue is discussed further in the main text in \S~\ref{prospects}.
\vspace{2ex}

\begin{acknowledgments}
We thank Luis Anchordoqui, Sidney Bludman, Sergio Bottai, Zoltan Fodor, 
Peter Gorham, Steen Hannestad, Michael Kachelriess, Sandor Katz, Chris Quigg, 
Christian Spiering, and Yvonne Wong for useful discussions and important information. 
Help in programming issues from Hui Fang is also kindly acknowledged. 
TJW thanks the CfCP at the U. Chicago, KIPAC at Stanford/SLAC, 
US DoE grant DE-FG05-85ER40226,
NASA-ATP grant 02-0000-0151, 
and Vanderbilt University's Discovery Grant program 
for sabbatical support during preparation of this paper.
\end{acknowledgments}


\end{document}